\documentclass[3p]{elsarticle}
\usepackage[titletoc,toc,title]{appendix}
\usepackage{euscript,amsmath,amssymb,amsfonts,graphicx,bm}
\usepackage{hyperref}
\hypersetup{colorlinks=true,linkcolor=blue,citecolor=red}
\usepackage{epstopdf}
\usepackage{enumerate}
\usepackage{mathrsfs}
\usepackage{xcolor}
\usepackage{tikz}
\usetikzlibrary{datavisualization.formats.functions}
\usepackage{algorithm}
\usepackage{enumerate}
\usepackage[noend]{algpseudocode}

\newcommand{\x}{\bm{x}}
\newcommand{\z}{\bm{z}}
\newcommand{\y}{\bm{y}}
\newcommand{\Y}{\bm{Y}}
\newcommand{\X}{\bm{X}}
\newcommand{\W}{\bm{W}}
\newcommand{\calU}{\mathcal{U}}
\newcommand{\calT}{\mathcal{T}}
\newcommand{\calA}{\mathcal{A}}
\newcommand{\calD}{\mathcal{D}}
\newcommand{\calL}{\mathcal{L}}
\newcommand{\calF}{\mathcal{F}}
\newcommand{\calB}{\mathcal{B}}
\newcommand{\calC}{\mathcal{C}}
\newcommand{\calO}{\mathcal{O}}
\newcommand{\calR}{\mathcal{R}}

\newcommand{\calJ}{\mathcal{J}}
\newcommand{\calJl}{\widetilde{\mathcal{J}}}
\newcommand{\calI}{\mathcal{I}}
\newcommand{\calM}{\mathcal{M}}
\newcommand{\calG}{\bm{\mathcal{G}}}
\newcommand{\calH}{\bm{\mathcal{H}}}
\newcommand{\Z}{\mathcal Z}
\newcommand{\R}{\mathbb R}
\newcommand{\Lh}{\widehat{L}}

\newcommand{\n}{\bm{n}}
\newcommand{\Ah}{\widehat{A}}

\newcommand{\prob}[1]{\mathbb{P}\left[#1\right]}
\newcommand{\expect}[1]{\mathbb{E}\left[#1\right]}
\newcommand{\inner}[1]{\left<#1\right>}
\newcommand{\bbT}{\mathbb{T}}
    \numberwithin{equation}{section}

\begin{document}

\begin{frontmatter}
\title{A numerical method for solving snapping out Brownian motion in 2D bounded domains}

\author{Ryan D. Schumm$^1$}
\ead{schumm@math.utah.edu}

\author{Paul C. Bressloff$^{1}$}
\ead{bressloff@math.utah.edu}

\address{$^1$Department of Mathematics, University of Utah, Salt Lake City, UT 84112 USA}

    \begin{abstract}
    
 Diffusion in heterogeneous media partitioned by semi-permeable interfaces has a wide range of applications in the physical and life sciences, including gas permeation in soils, diffusion magnetic resonance imaging (dMRI), drug delivery, thermal conduction in composite media, synaptic receptor trafficking, and intercellular gap junctions. At the single particle level, diffusion across a semi-permeable interface can be formulated in terms of so-called snapping out Brownian motion (SNOBM). The latter sews together successive rounds of reflected BM, each of which is restricted to one side of the interface. Each round of reflected BM is killed when the local time at the interface exceeds an independent, exponentially distributed random variable. (The local time specifies the amount of time a reflected Brownian particle spends in a neighborhood of the interface.) The particle then immediately resumes reflected BM on the same side or the other side of the interface according to a stochastic switch, and the process is iterated. In this paper, we develop a  Monte Carlo algorithm for simulating a two-dimensional version of SNOBM, which is used to solve a first passage time (FPT) problem for diffusion in a domain with semi-permeable partially absorbing traps. Our method combines a walk-on-spheres (WOS) method with an efficient algorithm for computing the boundary local time that uses a Skorokhod integral representation of the latter. We validate our algorithm by comparing the Monte Carlo estimates of the MFPT to 
    the exact solution for a single circular trap, and show that our simulations are consistent with asymptotic results obtained for the 2D narrow capture problem involving multiple small circular targets. We also discuss extensions to higher dimensions.

    \end{abstract}
    

\end{frontmatter}

    \section{Introduction}
    The mathematical analysis of single-particle diffusion through a semi-permeable interface is important for our understanding of transport phenomena in physical and biological systems. 
 Such processes include molecular transport through lipid bilayers \cite{bio-app-1,bio-app-2,bressloff-book,bio-app-3}, the dynamics of gap junctions \cite{Brink85,Connors04,Bressloff16}, thermal conduction in composite media \cite{Grossel98,deMonte00,Lu05}, diffusion magnetic resonance imaging (dMRI) \cite{Tanner78,Callaghan92,Coy94,Grebenkov14} and drug delivery \cite{Pontrelli07,Todo13,Farago18}. Furthermore, it was recently shown that the trafficking of neurotransmitter receptor proteins in the postsynaptic membrane of neurons can be mathematically formulated in terms of a reaction-diffusion system involving semipermeable membranes which separate the bulk of the neuronal membrane from the synaptic regions \cite{bressloff-synapse}.  A mathematical understanding of this phenomenon provides insights into how synaptic strengths are modulated during learning and memory \cite{schumm-synapse}. 
    
Population level models for transport through semi-permeable membranes can be formulated using the Kedem-Katchalsky equations, which were originally derived
    using arguments from statistical thermodynamics \cite{kk-1, kk-2, kk-3}.  Alternatively, these processes can be described at the level of single-particle diffusion, which allows for 
    the utilization of tools from stochastic analysis and Monte Carlo methods. It has been shown that one-dimensional (1D) diffusion through a semi-permeable interface at the origin
  is equivalent to a process called snapping out Brownian motion (SNOBM). The latter links together a sequence of reflected Brownian motions (BMs) that are killed at the semipermeable interface and then reset on either side of the interface as determined by a stochastic switch \cite{lejay,Lejay18}. Each round of reflected BM is killed when its local time at $x=0^{\pm}$ exceeds an exponentially distributed random variable. (The local time at $x=0^+$ ($x=0^-$) is a Brownian functional that specifies the amount of time a positively (negatively) reflected Brownian particle spends in contact with the right-hand (left-hand) side of the interface \cite{lt-1,lt-2,lt-3}.) 
  
 Recently, 1D SNOBM has been reformulated in terms of a renewal equation that relates the full probability density of particle position to the probability densities of partially reflected BMs on either side of the interface \cite{ bressloff-snob-1d}.  (In Ref. \cite{lejay} a corresponding backward equation was derived using the theory of semigroups and resolvent operators.)  It can be shown that the solution of the renewal equation satisfies the single-particle diffusion equation with boundary conditions imposed on the semipermeable interface that are equivalent to that of the Kedem-Katchalsky equations. The renewal theory of SNOBM has also been extended to bounded domains and higher spatial dimensions, and to include non-Markovian mechanisms for killing each round of reflecting BM \cite{bressloff-snob-high-d,3d-narrow-capture}. The latter leads to a time-dependent permeability that tends to be heavy-tailed. Formulating interfacial diffusion in terms of SNOBM thus provides a general probabilistic framework for modeling semi-permeable membranes. Another important feature of SNOBM is that it is a stochastic process that generates exact sample paths of BM in the presence of a semi-permeable interface. This implies that developing an efficient numerical scheme for simulating SNOBM in multiple dimensions could be used to obtain approximate solutions of non-trivial boundary value problems (BVPs) in the presence of  semi-permeable interfaces. The construction of such a scheme for two-dimensional (2D) domains is the main goal of the current paper. 
 A computational method for finding solutions to the 1-D single-particle diffusion equation in the presence of one or more 
 semi-permeable interfaces has been developed in terms of underdamped Langevin equations \cite{Farago18,Farago20}.  In this method, the particle trajectory is simulated 
 by sampling a sequence of Gaussian random numbers and numerically solving the Langevin equations using the GJF integrator. When the particle encounters a semi-permeable interface,
 the particle is reflected with a fixed probability that depends on the mass of the particle and permeability of the interface.  
 The reflection probability is derived by assuming
 the particle ensemble is in thermal equilibrium and applying Fick's first law. 
 However, this is distinct from SNOBM, which is an exact single-particle realization of diffusion through an interface in the over-damped limit.  Additionally, our SNOBM based
 numerical method uses computationally efficient kinetic Monte Carlo techniques and requires no numerical integration.

In order to construct our numerical algorithm we focus on the particular problem of diffusion in a bounded 2D domain $\Omega \subset \R^2$ containing one or more partially absorbing traps $\calU_j\subset \Omega$, $j=1,\ldots,N$. The boundary $\partial \calU_j$ of the $j$-th trap is taken to be a closed 1D semi-permeable interface with constant permeability $\kappa_j\in(0,\infty)$ and directional bias $\alpha_j\in(0,1)$. Thermodynamically speaking, the latter could be interpreted as a discontinuity in the chemical potential across the interface.
Whenever a particle enters a trap, it can be absorbed at a constant Poisson rate $\gamma$. It follows that natural quantities of interest include the splitting probabilities and unconditional mean first passage time (MFPT) for absorption by the traps. Each of these quantities satisfies a non-trivial BVP. 
In this paper we use the stochastic differential equations (SDEs) of SNOBM to build Monte Carlo numerical methods for simulating the reaction-diffusion process and thus finding numerical solutions of the corresponding BVPs. Effective simulation of SNOBM is highly dependent on accurate computations of the boundary local times that are used to determine when each round of reflecting BM is killed. (Similar issues arise in encounter-based models of partially absorbing reactive surfaces \cite{grebenkov-1, grebenkov-2, bressloff-bf}.)  Accurately and efficiently computing boundary local times can be numerically challenging and depends on the mathematical representation of the local time \cite{sylvain, zhou,grebenkov-1, grebenkov-2}.  

The numerical scheme presented in this paper consists of two major elements. First, we use a walk-on-spheres (WOS) algorithm to compute the time for each excursion from a point in the interior of the domain to enter a small $\epsilon$-neighborhood of the boundary $\partial \Omega \cup_{j=1}^N\partial \calU_j$. The WOS algorithm involves simulating the trajectory of a Brownian particle by treating the dynamics as a sequence of diffusions inside of spheres (disks in 2D) with totally absorbing boundaries.
    The position of the particle is updated by randomly sampling a point on the surface of the sphere and then repeating the process until the particle enters the boundary layer. When the particle is sufficiently far from any boundaries, this method of simulation provides exact results and is more computationally efficient than the standard Euler-Maruyama method. The WOS algorithm was originally introduced by M\"uller in \cite{wos} to solve Dirichlet BVPs.  For this application, the simulation is terminated once the particle enters the boundary layer and is projected
    onto the nearest boundary point.  In order to compute the solution to the Laplace BVP, one only needs the boundary point at the termination time.  On the other hand, to determine quantities such as the MFPT for more complicated reaction-diffusion
    processes, one also needs to compute the time elapsed between successive WOS iterations. 
We implement this using an exact analytical formula for the survival probability of diffusion in a totally absorbing disk similar to Ref. \cite{kinetic-monte-carlo}. The second component of our numerical scheme is an accurate method for computing the local time when the Brownian particle is in an $\epsilon$-neighborhood of the boundary, which is based on the Skorokhod integral representation of the boundary local time \cite{elton}. This allows us to calculate the local times with much higher accuracy than previous algorithms.  Additionally, the Skorokhod integral representation requires limited data generated from particle-boundary collisions, which simplifies simulating the particle trajectory near a boundary and lowers computation times.

The structure of the paper is as follows.  In section 2, we describe the two complementary approaches to calculating the MFPT and splitting probabilities associated with the 2D reaction-diffusion process involving 
    multiple partially absorbing semi-permeable traps.  First, we derive BVPs using the forward and backward reaction-diffusion equations and find the exact solution to the MFPT BVP 
    for a single circular trap.  Next, we formulate 2D SNOBM in terms of a sequence of killed reflected Brownian
    motions and derive the associated SDEs and random killing times.
    In section 3, we present our stochastic simulation algorithm that solves the SDEs of SNOBM and generates Monte Carlo estimates of the MFPT and splitting probabilities. 
    In section 4, we perform a number of numerical tests to evaluate the accuracy and convergence of the SNOBM Monte Carlo simulations.  First,  we compare our local time algorithm to one developed in Ref. \cite{zhou} by solving for the MFPT in a 
    unit disk with a partially reactive boundary. Second, we compare the Monte Carlo estimates of the MFPT to the exact solution of the BVP for a single circular trap. Finally, we consider a more complicated configuration consisting of three circular traps within the unit disk. In this case we cannot obtain an exact analytical solution of the BVPs. Therefore, we take the trap radii to be much smaller than unity, which allows us to obtain approximations of the MFPT and splitting probabilities using Green's function and matched asymptotic methods. Such methods are used widely to solve so-called narrow capture problems \cite{Ward15,narrow-capture-dirichlet,narrow-capture-robin}, which we adapt to include the effects of semi-permeable interfaces. (The details are presented in the appendix.) We show that our asymptotic results are in good agreement with the corresponding numerical simulations for sufficiently small traps, and explore how errors increase with the size of the traps. 
    
   \section{Diffusion with semi-permeable partially absorbing traps}
    \begin{figure}[b!]
        \centering
        \includegraphics[width=12cm]{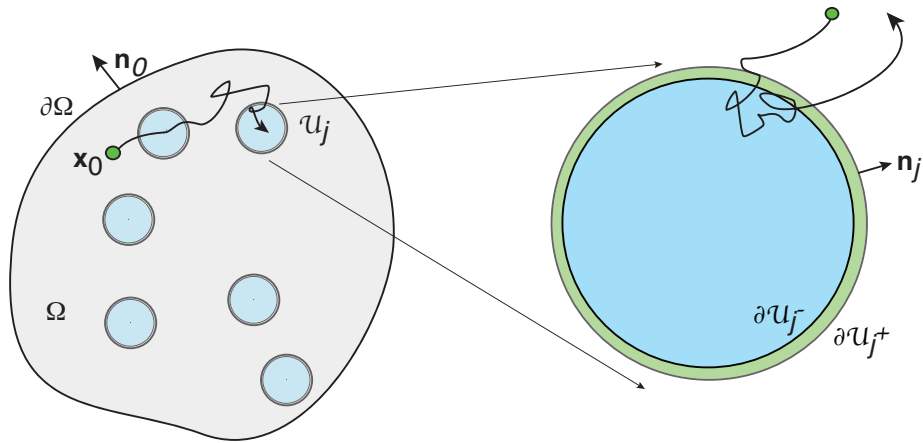} 
        \caption{A particle diffuses in a bounded domain $\Omega\subset \R^2 $ containing $N$ partially absorbing traps $\calU_j$, $j=1,\ldots,N$. Whenever the particle is within the trap domain $\calU_j$, it can be absorbed at a rate $\gamma$.The outward unit normals of $\partial \Omega$ and $\partial \calU_j$ are denoted by $\n_0$ and $\n_j$, respectively. [Insert: The boundary $\partial \calU_j$ of the $j$th trap acts as a semipermeable interface.]
                }
        \label{fig1}
    \end{figure}
    Consider a Brownian particle with position $\X_t$ at time $t$ diffusing in a bounded domain $\Omega\subset\mathbb{R}^2$ containing $N$ traps $\calU_j\subset\Omega$, 
    $j=1,\ldots,N$ where $\partial\Omega$ is a smooth totally reflecting boundary, see Fig. \ref{fig1}. Each trap boundary $\partial\calU_j$ acts as semi-permeable interface with permeability $\kappa_j$ 
    and directional bias $\alpha_j\in [0, 1]$. The particle has a space dependent diffusivity given by
    \begin{gather}
        D(\x) = 
        \begin{cases}
            D_0, \ \x\in\Omega\backslash\calU,\\
            D_j, \ \x\in\calU_j, \ j=1,\ldots,N,
        \end{cases}
    \end{gather}
    where $D_0$ and $D_j$ are positive constants.
    Let $\calU_a = \cup_{j=1}^N\calU_j$ and the indicator function be denoted by
    \begin{gather}
        I_{\calU_j}(\x) = 
        \begin{cases}
            1, \ \x \in \calU_j\\
            0, \ \x \notin \calU_j
        \end{cases}.
    \end{gather}
    Absorption occurs when there exists a $j=1,\ldots,N$ such that $A^{(j)}_t \geq \Ah_j$ where
    \begin{align}
        \label{occupation}
        A^{(j)}_t = \int_0^t I_{\calU_j}\left(\X_\tau\right)d\tau
    \end{align}
    is a Brownian functional called the occupation time that tracks the amount of time the particle has spent in the trap region $\calU_j$ and $\Ah_j$ is an independent 
    random time with 
    \begin{align}
        \prob{\Ah_j < a} = 1 - e^{-\gamma a}, \ a \geq 0.
    \end{align}
    Define the first passage time as  
    \begin{align}
        \calT = \inf\left\{t > 0: \exists j \ \middle| \ A^{(j)}_t\geq\Ah_j\right\},
        \label{calT}
    \end{align}
   The unconditional MFPT and the splitting probabilities are then
    \begin{subequations}
        \begin{align}
            &T(\x) = \expect{\calT\middle|\X_0 = \x}, \label{mfpt-def}\\
            &\pi_j(\x) = \prob{\X_\calT\in\calU_j\middle|\X_0=\x}, \ j = 1,\ldots,N, \label{sp-def}
        \end{align}
    \end{subequations}
    respectively. One can use forward and backward reaction-diffusion equations to derive boundary value problems for the MFPT and splitting probabilities along the lines
    of \cite{3d-narrow-capture}.  Alternatively, one can formulate the reaction-diffusion process in terms of SDEs, which can be solved using Monte Carlo methods. 
    Both approaches are described below.

    \subsection{Derivation of the boundary value problems}
    The probability density of the diffusion process denoted by
    \begin{gather}
        p(\x, t|\y)d\x = \prob{\X_t\in(\x, \x + d\x)\middle| \X_0=\y\in\Omega}
    \end{gather}
    satisfies the following system of forward reaction diffusion equations \cite{bressloff-snob-high-d}:
    \begin{subequations}
        \label{forward}
        \begin{align}
            &\frac{\partial}{\partial t}p(\x, t|\y) = D_0\nabla^2_{\x} p(\x, t|\y), \ \x\in\Omega\backslash\calU_a,\\
            &\nabla_{\x} p(\x, t|\y)\cdot\n_0\equiv J(\x,t|\y) = 0, \ \x\in\partial\Omega, \label{forward-exterior}\\
            &\frac{\partial}{\partial t}p(\x, t|\y) = D_j\nabla^2_{\x}p(\x, t|\y) - \gamma p(\x, t|\y), \ \x\in\calU_j,\\
            &D_0\nabla_{\x} p(\x^+, t|\y)\cdot\n_j = D_j\nabla_{\x} p(\x^-, t|\y)\cdot\n_j \equiv J_j(\x, t|\y), \ \x\in\partial\calU_j, \label{forward-bc1}\\
            &J_j(\x, t|\y) = \kappa_j\left[(1 - \alpha_j)p(\x^+, t|\y) - \alpha_jp(\x^-, t|\y)\right], \ \x\in\partial\calU_j ,\label{forward-bc2}
        \end{align}
    \end{subequations}
    where $\n_0$ and $\n_j$ are the outward pointing normal vectors to $\partial\Omega$ and $\calU_j$ respectively, see Fig. \ref{fig1}.
    The probability density function of $\calT$ is given by 
    \begin{align}
        \prob{\calT\in(t, t + dt)\middle|\X_0=\y} = -\frac{\partial Q(\y, t)}{\partial t}dt,
    \end{align}
    where
    \begin{align}
        Q(\y, t) = \prob{\calT > t\middle|\X_0=\y} = \int_{\Omega}p(\x, t|\y)d\x
    \end{align}
    is the survival probability of the reaction-diffusion process.  Using equation (\ref{forward}) and the divergence theorem, we can write
    \begin{align}
        -\frac{\partial Q(\y, t)}{\partial t} &= -\int_{\Omega\backslash\calU_a}D_0\nabla^2p(\x, t|\y)d\x - \sum_{j=1}^N\int_{\calU_j}\left[D_j\nabla^2p(\x,t|\y) - \gamma p(\x, t|\y)\right]d\x\nonumber\\
        &= \sum_{j=1}^N\left[\int_{\partial\calU^+_j}D_0\nabla p(\x, t|\y)\cdot\n_1d\x - \int_{\partial\calU^-_j}D_j\nabla p(\x, t|\y)\cdot\n_1d\x + \gamma\int_{\calU_j}p(\x, t|\y)d\x\right]\nonumber\\
        &= \gamma\sum_{j=1}^N\int_{\calU_j}p(\x, t|\y)d\x \equiv \sum_{j=1}^N\calJ_j(\y, t).
    \end{align}
    Therefore, the unconditional MFPT can be written as
    \begin{align}
        \label{flux-to-mfpt}
        T(\y) = \int_0^\infty t\sum_{j=1}^N\calJ_j(\y, t)dt 
        = \sum_{j=1}^N\int_0^\infty-\left.\frac{\partial}{\partial s}\left[e^{-st}\calJ_j(\y, t)\right]\right|_{s=0}dt 
        = -\sum_{j=1}^N\left.\frac{\partial}{\partial s}\calJl_j(\y, s)\right|_{s=0},
    \end{align}
    where $\calJl(\x, s)$ is the Laplace transform of the probability flux into the trap region. The splitting probabilities can be written as
    \begin{align}
        \label{flux-to-sp}
        \pi_j(\y) = \int_0^\infty\calJ_j(\y, t)dt = \calJl_j(\y, 0).
    \end{align}
    Using equations (\ref{flux-to-mfpt}) and (\ref{flux-to-sp}) along with the backward reaction-diffusion equation derived bellow, we can obtain BVPs for the MFPT and
    splitting probabilities.
    
    The generator of the diffusion process is $\calL = \nabla^2$.  Let $\calD(\calL)\subset L^2(\Omega)$ be the operator domain with associated boundary conditions (\ref{forward-exterior}),
    (\ref{forward-bc1}), and (\ref{forward-bc2}). We have that $\calL = \calL^\dag$ but the boundary conditions of the adjoint domain $\calD(\calL^\dag)$ need to be calculated 
    explicitly. Observe that for $f \in \calD(\calL)$ and $g \in \calD(\calL^\dag)$, we have that 
    \begin{align}
        \label{inner-prod}
        \inner{\calL f, g} = \int_\Omega g\nabla^2fd\x = \int_{\Omega\backslash\calU_a} g\nabla^2fd\x + \sum_{j=1}^N\int_{\calU_j} g\nabla^2fd\x
    \end{align}
    Using Green's first identity and applying the exterior boundary condition (\ref{forward-exterior}) to $f$, the first integral in (\ref{inner-prod}) can be written as
    \begin{align}
        \label{green-identity-outer}
        \int_{\Omega\backslash\calU_a} g\nabla^2fd\x &= -\int_{\Omega\backslash\calU_a}\nabla f\cdot\nabla g d\x 
            - \sum_{j=1}^N\int_{\partial\calU_j^+}g\nabla f\cdot\n_j d\x \nonumber\\
        &= \int_{\Omega\backslash\calU_a}f\nabla^2gd\x - \int_{\partial\Omega}f\nabla g\cdot\n_0 d\x
            - \sum_{j=1}^N\int_{\partial\calU_j^+}\left[g\nabla f\cdot\n_j - f\nabla g\cdot\n_j\right] d\x.
    \end{align}
    Similarly, we have that
    \begin{align}
        \label{green-identity-inner}
        \int_{\calU_j} g\nabla^2fd\x = \int_{\calU_j}f\nabla^2gd\x + \int_{\partial\calU_j^-}\left[g\nabla f\cdot\n_j - f\nabla g\cdot\n_j\right]d\x .
    \end{align}
    Substituting equations (\ref{green-identity-outer}) and (\ref{green-identity-inner}) into (\ref{inner-prod}) yields
    \begin{align}
        \label{inner-prod-2}
        \inner{\calL f, g} = &\inner{f, \calL^\dag g} - \int_{\partial\Omega}f\nabla g\cdot\n_0 d\x
            + \sum_{j=1}^N\left (\int_{\partial\calU_j^-} - \int_{\partial\calU_j^+}\right )\left[g\nabla f\cdot\n_j - f\nabla g\cdot\n_j\right] d\x .    \end{align}
    The boundary integrals in equation (\ref{inner-prod-2}) must sum to zero for all $f\in\calD(\calL)$ and $g\in\calD(\calL^\dag)$.  Therefore,
    we can impose the adjoint boundary conditions
    \begin{subequations}
        \begin{align}
            &\nabla g\cdot\n_0 = 0, \ \x \in \partial\Omega,\\
            &g\nabla f\cdot\n_j - f\nabla g\cdot\n_j = 0, \ \x\in\partial\calU_j^\pm\label{adjoint-bc}
        \end{align}
    \end{subequations}
    Take $f(\x) = p(\x, t|\y)$ and $g(\x) = p(\z, t|\x)$ and set 
    \begin{align}
        &\calI_j(\z, t| \x^\pm) = D(\x^\pm)\nabla_{\x} p(\z, t|\x^\pm)\cdot\n_j, \ \x^\pm\in\partial\calU_j^\pm
    \end{align}
    From (\ref{adjoint-bc}), we have that
    \begin{align}
       p(\z, t|\x^+)J_j(\x, t|\y) - p(\x^+, t|\y)\calI_j(\z, t|\x^+) =0=
      p(\z, t|\x^-)J_j(\x, t|\y) - p(\x^-, t|\y)\calI_j(\z, t|\x^-).
    \end{align}
    Therefore, equation (\ref{forward-bc2}) implies that
    \begin{align}
        \calI_j(\z, t|\x^-)p(\x^-, t|\y) - \calI_j(\z, t|\x^+)p(\x^+, t|\y) = \kappa_j\alpha_j\left[p(\z, t|\x^+) - p(\z, t|\x^-)\right]p(\x^-, t|\y)\nonumber\\
            \quad - \kappa_j(1 - \alpha_j)\left[p(\z, t|\x^+) - p(\z, t|\x^-)\right]p(\x^+, t|\y).
    \end{align}
    Equating coefficients of $p(\x^+, t|\y)$ and $p(\x^-, t|\y)$ yields
    \begin{subequations}
        \begin{align}
            &\calI_j(\z, t|\x^+) = \kappa_j(1 - \alpha_j)\left[p(\z, t|\x^+) - p(\z, t|\x^-)\right],\\
            &\calI_j(\z, t|\x^-) = \kappa_j\alpha_j\left[p(\z, t|\x^+) - p(\z, t|\x^-)\right]
        \end{align}
    \end{subequations}
    which implies that $\alpha_j\calI_j(\z, t|\x^+) = (1 - \alpha_j)\calI_j(\z, t|\x^-)$. Therefore, the backward equations can be written as
    \begin{subequations}
        \label{backward}
        \begin{align}
            &\frac{\partial}{\partial t}p(\z, t|\x) = D_0\nabla^2_{\x} p(\z, t|\x), \ \x\in\Omega\backslash\calU_a,\\
            &\nabla_{\x} p(\z, t|\x)\cdot\n_0 = 0, \ \x\in\partial\Omega, \label{backward-exterior}\\
            &\frac{\partial}{\partial t}p(\z, t|\x) = D_j\nabla^2_{\x}p(\z, t|\x) - \gamma p(\z, t|\x), \ \x\in\calU_j,\\
            &\alpha_jD_0\nabla_{\x} p(\z, t|\x^+)\cdot\n_j = (1 - \alpha_j)D_j\nabla_{\x} p(\z, t|\x^-)\cdot\n_j \equiv J_j^\dag(\z, t|\x), \ \x\in\partial\calU_j, 
                \label{backward-bc1}\\
            &J_j^\dag(\z, t|\x) = \kappa_j\alpha_j(1 - \alpha_j)\left[p(\z, t|\x^+) - p(\z, t|\x^-)\right], \ \x\in\partial\calU_j \label{backward-bc2}.
        \end{align}
    \end{subequations}
    Integrating (\ref{backward}) with respect to $\z$ over $\calU_j$, Laplace transforming, and applying the initial condition 
    \begin{align}
        \calJ_j(\x, 0) = \gamma\int_{\calU_j}\delta(\z - \x)d\z = \gamma I_{\calU_j}(\x),
    \end{align}
    one finds that 
    \begin{subequations}
        \label{flux-bvp}
        \begin{align}
            &D_0\nabla^2\calJl_j(\x, s) - s\calJl_j(\x, s) = 0, \ \x\in\Omega\backslash\calU_a,\\
            &\nabla \calJl_j(\x, s)\cdot\n_0 = 0, \ \x\in\partial\Omega, \label{flux-exterior}\\
            &D_k\nabla^2\calJl_j(\x, s) - (s + \gamma)\calJl_j(\x, s) = -\gamma\delta_{jk}, \ \x\in\calU_k,\\
            &\alpha_kD_0\nabla\calJl_j(\x^+, s)\cdot\n_k = (1 - \alpha_k)D_k\nabla\calJl_j(\x^-, s)\cdot\n_k \equiv \widetilde{J}_{jk}(\x, s), 
                \ \x\in\partial\calU_k, \label{flux-bc1}\\
            &\widetilde{J}_{jk}(\x, s) = \kappa_k\alpha_k(1 - \alpha_k)\left[\calJl_j(\x^+, s) - \calJl_j(\x^-, s)\right], \ \x\in\partial\calU_k \label{flux-bc2}
        \end{align}
    \end{subequations}
    Letting $s\to 0$, and using the fact that 
    \begin{align}
        \lim_{s\to 0}s\calJl_j(x, s) = \lim_{t\to\infty}\calJ_j(\x, t) = 0
    \end{align}
    yields the BVP
    \begin{subequations}
        \label{sp-bvp}
        \begin{align}
            &\nabla^2\pi_k(\x) = 0, \ \x\in\Omega\backslash\calU_a,\\
            &\nabla\pi_k(\x)\cdot\n_0 = 0, \ \x\in\partial\Omega,\\
            &\nabla^2\pi_k(\x) - \frac{\gamma}{D_j}\pi_k(\x) = -\frac{\gamma}{D_j}\delta_{j, k}, \ \x\in\calU_j,\\
            &\alpha_jD_0\nabla\pi_k(\x^+)\cdot\n_j = (1 - \alpha_j)D_j\nabla\pi_k(\x^-)\cdot\n_j \equiv \mathcal{P}_{k, j}(\x), \ \x\in\partial\calU_j,\\
            &\mathcal{P}_{k, j}(\x) = \kappa_j\alpha_j(1 - \alpha_j)\left[\pi_k(\x^+) - \pi_k(\x^-)\right] , \ \x \in \partial\calU_j.
        \end{align}
    \end{subequations}
    Summing equation (\ref{flux-bvp}) over $j=1,\ldots,N$, differentiating with respect to $s$, and letting $s\to 0$ yields
    \begin{subequations}
        \label{mfpt-bvp}
        \begin{align}
            &\nabla^2T(\x) = -\frac{1}{D_0}, \ \x\in\Omega\backslash\calU_a,\\
            &\nabla T(\x)\cdot\n_0 = 0, \ \x\in\partial\Omega,\\
            &\nabla^2T(\x) - \frac{\gamma}{D_j} T(\x) = -\frac{1}{D_j}, \ \x\in\calU_j,\\
            &\alpha_jD_0\nabla T(\x^+)\cdot\n_j = (1 - \alpha_j)D_j\nabla T(\x^-)\cdot\n_j \equiv \mathcal{M}_{j}(\x), \ \x\in\partial\calU_j,\\
            &\mathcal{M}_j(\x) = \kappa_j\alpha_j(1 - \alpha_j)\left[ T(\x^+) -  T(\x^-)\right], \ \x \in \partial\calU_j.
        \end{align}
    \end{subequations}
    An exact solution to BVP (\ref{mfpt-bvp}) can be obtained when $N=1$ and $\Omega$ and $\calU$ are concentric disks.  For $N > 2$, one can obtain 
    approximate solutions to both (\ref{mfpt-bvp}) and (\ref{sp-bvp}) using matched asymptotic analysis and Green's function methods (see appendix).
    
 Various limiting cases have been analyzed elsewhere. For example, in the limit $\kappa_j \to \infty$ with $\alpha_j=1/2$, the particle can freely diffuse into the trap $\calU_j$, which acts as partially absorbing interior with absorption rate $\gamma$.
    This problem was analyzed in \cite{narrow-capture-reset}.  On the other hand, if $\gamma\to\infty$ and the $\kappa_j$ remain finite, then the boundaries $\partial\calU_j$ act as partially reactive surfaces and the boundary conditions (\ref{forward-bc1}) and (\ref{forward-bc1}) are replaced with the Robin boundary conditions \cite{narrow-capture-robin}
    \begin{align}
        D\nabla p(\x, t|\y)\cdot\n_j = \kappa_jp(\x, t|\y), \ \x\in\partial\calU_j, \ j=1,\ldots,N.
    \end{align}
   Finally, if both $\kappa_j\to\infty$ and $\gamma\to\infty$, then the trap boundaries are totally absorbing with
    Dirichlet boundary conditions \cite{Ward15,narrow-capture-dirichlet, narrow-capture-flux}
    \begin{align}
        p(\x, t|\y) = 0, \ \x\in\calU_j, \ j=1,\ldots,N.
    \end{align}
        
    \subsection{A single circular trap in the unit disk}
    Let $\Omega$ be the unit disk containing a single circular trap $\calU_1$ centered at the origin with radius $R < 1$.  
    If we let $r = \|\x\|$ and $D_0 = D_1$, then the BVP (\ref{mfpt-bvp}) becomes
    \begin{subequations}
        \label{mfpt-spehre}
        \begin{align}
            &\frac{\partial^2T}{\partial r^2} + \frac{1}{r}\frac{\partial T}{\partial r} = -\frac{1}{D_0}, \ r \in (R, 1),\label{ode1}\\
            &\partial_rT(1)  = 0,\label{bc-outter}\\ 
            &\frac{\partial^2T}{\partial r^2} + \frac{1}{r}\frac{\partial T}{\partial r} - \frac{\gamma}{D_0}T = -\frac{1}{D_0}, \ r\in(0, R),\label{ode2}\\ 
            &\alpha\partial_r T(R^+) = (1 - \alpha)\partial_rT(R^-) = \frac{\kappa\alpha(1 - \alpha)}{D_0}\left[ T(R^+) -  T(R^-)\right]\label{bc-inner}.
        \end{align}
    \end{subequations}
    The general solution to equations (\ref{ode1}) and (\ref{ode2}) is given by
    \begin{gather}
        \label{sphere-general}
        T(r) = 
        \begin{cases}
            AI_0\left(\sqrt{\gamma / D_0}r\right) + 1/\gamma, \ r \in [0, R^-],\\ 
            -r^2/4D_0 + B\ln(r) + C, \ r \in [R^+, 1],
        \end{cases}
    \end{gather}
    where $A$, $B$, and $C$ are constants.  Substituting equation (\ref{sphere-general}) into equations (\ref{bc-outter}) 
    and (\ref{bc-inner}) and solving for $A$, $B$, and $C$ yields
    \begin{align}
        \label{single-target-solution}
        T(\x) = \frac{1}{2\gamma D_0\kappa R(1 - \alpha)}
            \left[\eta(\x) + \alpha\sqrt{\gamma D_0}\kappa(1 - R^2)\frac{I_0\left(\sqrt{\gamma / D_0}R\right)}{I_1\left(\sqrt{\gamma / D_0}R\right)}\right], \ \|\x\|>R,
    \end{align}
    where
    \begin{align}
        &\eta(\x) = \kappa\gamma R(1 - \alpha)\left[\ln\left(\frac{\|\x\|}{R}\right) - \frac{\|\x\|^2}{2} + \frac{R^2}{2} + \frac{2D_0}{\gamma}\right] + \gamma D_0(1 - R^2).
    \end{align}

    \subsection{Snapping out Brownian motion}
    An alternative approach to solving BVPs for the MFPT and splitting probabilities is to reformulate the reaction-diffusion process 
    in terms of the SDEs and random killing times of multi-dimensional SNOBM. This representation allows us to develop a stochastic simulation algorithm that computes
    Monte Carlo estimates of the MFPT and splitting probabilities.
    
   Let $\X_t$ denote the position of the particle at time $t$.
   The dynamics of SNOBM can be described in terms of a sequence of killed reflecting Brownian motions \cite{lejay, bressloff-snob-1d, bressloff-snob-high-d} 
    in either $\Omega\backslash\calU_a$ or $\calU_j$, $j=1,\ldots,N$. Let $\bbT_n$ denote the time of the $n^{\rm th}$ killing (with $\bbT_0 = 0$), and suppose that this occurs at position $\X_{\bbT_n} = \z_n \in \partial\calU_j$. Immediately after the killing event, the position of the particle is taken to be
    \begin{align}
        \label{killing-position}
        \X_{\bbT_n} = \mathcal{B}_j^{(n)}\z^-_n + \left(1 - \mathcal{B}_j^{(n)}\right)\z^+_n,
    \end{align}
    where $\mathcal{B}^{(n)}_j\sim\text{Ber}(\alpha_j)$ is an independent Bernoulli random variable. That is, the particle executes the next round of reflecting BM in $\Omega\backslash\calU_a$ with probability $1 - \alpha_j$ and in $\calU_j$ with probability $\alpha_j$.  Suppose that $\X_t\in\Omega\backslash\calU_a$ for $t \in (\bbT_{n}, \bbT_{n+1})$, that is, $ \X_{\bbT_n}=z_n^+$, and introduce the boundary local times 
    \begin{align}
        \label{local-time-main}
        L_t^{(k)}(\bbT_n) = \lim_{h \to 0}\frac{D}{h}\int_{\bbT_n}^t\Theta\left(h - \text{dist}(\X_\tau, \partial\calU_k^+)\right)d\tau,\quad k=1,\ldots,N,
    \end{align}
    and
 \begin{align}
        \label{local-time-b}
        L_t^{(b)}(\bbT_n) = \lim_{h \to 0}\frac{D}{h}\int_{\bbT_n}^t\Theta\left(h - \text{dist}(\X_\tau, \partial\Omega)\right)d\tau,
    \end{align}
   where $\Theta$ denotes the Heaviside function. The boundary local times $L^{(j)}_t(\bbT_n)$ and $L^{(b)}_t(\bbT_n)$ are a set of Brownian functionals that track the amount of the time the particle is in contact with the boundaries 
    $\partial\calU_j$ and $\partial \Omega$ over the time interval $[\bbT_n,t]$. It can be proven that the local times exist, and are continuous, positive increasing functions of time \cite{lt-1, lt-2, lt-3}.
  The SDE for $\X_t$, $t \in (\bbT_{n}, \bbT_{n+1})$, is given by the so-called Skorokhod equation for reflecting BM in the bounded domain $\Omega$ containing $N$ partially reactive surfaces 
    $\partial\calU_j$:
    \begin{align}
        \label{sde-outside}
        d\X_t = \sqrt{2D}d\W_t + \sum_{j=1}^N\n_j(\X_t)dL^{(j)}_t(\bbT_n) - \n_0(\X_t)dL^{(b)}_t(\bbT_n), \ \X_t \in \Omega\backslash\calU_a,\quad t \in (\bbT_{n}, \bbT_{n+1}).
    \end{align}
    Formally speaking, 
    \begin{equation}
    dL^{(j)}_t(\bbT_n)=\int_{\partial \calU_j}\delta(\X_t-\z)d\z,\quad  dL^{(b)}_t(\bbT_n)=\int_{\partial \Omega}\delta(\X_t-\z)d\z,
    \end{equation}
so that each time the particle hits a boundary it is given an impulsive kick back into the domain in a direction perpendicular to the boundary. 
   The time of the next killing is then determined by the condition
    \begin{align}
        \bbT_{n+1} = \inf\left\{t > \bbT_{n}: \exists k \ \middle| \ L^{(k}_t(\bbT_n)\geq\Lh_k\right\}, 
    \end{align}
and $\Lh_k$ is an independent randomly generated local time threshold with 
    \begin{align}
       \prob{\Lh_k < \ell} = 1 - e^{-\kappa_k\ell/2}, \ \ell \geq 0.
    \end{align}
On the other hand, if $\X_{\bbT_n} = \z_n^-$ then the next round of reflecting BM takes place in the domain $\calU_j$ with a single 
    partially reactive surface $\partial\calU_j$. The corresponding SDE is
    \begin{align}
        \label{sde-inside}
        d\X_t = \sqrt{2D}d\W_t - \n_j(\X_t)dL_t^{(j)}(\bbT_n), \ \X_t\in\calU_j, \quad t\in (\bbT_n,\bbT_{n+1}),
    \end{align}
    with 
    \begin{align}
        \label{local-time-j}
        L_t^{(j)}(\bbT_n) = \lim_{h \to 0}\frac{D}{h}\int_{\bbT_n}^t\Theta\left(h - \text{dist}(\X_\tau, \partial\calU_k^-)\right)d\tau
    \end{align}
so that
    \begin{align}
        \bbT_{n+1} = \inf\left\{t > \bbT_{n}:\exists j \ \middle| \ L^{(j)}_t(\bbT_n)\geq\Lh_j\right\}.
    \end{align} 
       
    In summary, the dynamics of SNOBM consists of sewing together successive rounds of reflecting BM, each of which evolves according to the SDE (\ref{sde-outside}) or (\ref{sde-inside}). Each round is killed when the local time at one of the accessible trap boundaries exceeds an exponentially distributed random threshold. (The threshold is independently generated each round.) Following each round of killing, a biased coin is thrown to determine which side of the interface the next round occurs. The sequence of reflected Brownian motions is permanently terminated when the amount of time spent in one of the traps exceeds its corresponding occupation time threshold, which occurs at the time $\calT$, see equation (\ref{calT}).

    \section{Stochastic simulation algorithm}
    In this section, we present a Monte Carlo algorithm that solves the SDEs of SNOBM and calculates the MFPT and splitting probabilities.  We first use a walk-on-spheres (WOS) method to calculate the time of each excursion from the interior of the domain to an $\epsilon$-neighborhood of the boundary. To speed up the algorithm, we use the exact solution for the survival probability in the unit disk to compute the WOS time increments. We then utilize a Skorokhod integral representation \cite{elton} to compute the boundary local times.
        
    \subsection{The walk-on-spheres method}
\begin{figure}[t!]
        \centering
        \includegraphics[width=14cm]{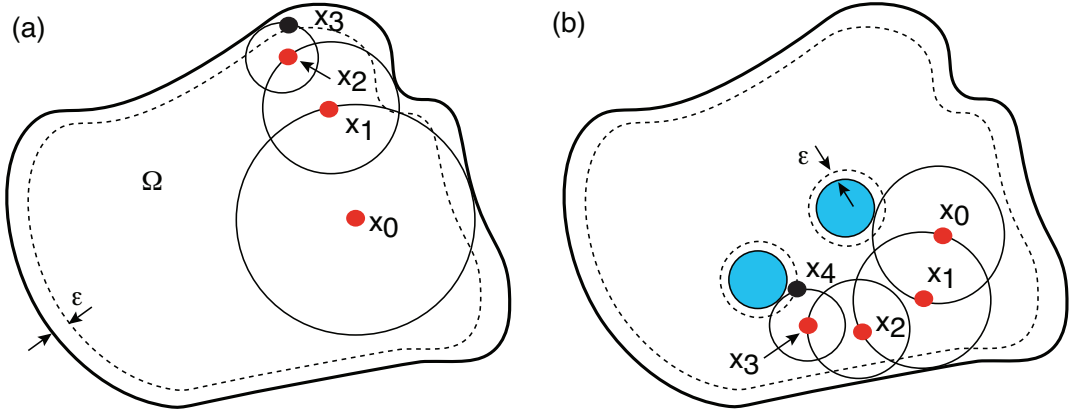} 
        \caption{Illustration of a single run of the walk-on-spheres algorithm for diffusion in a bounded domain 
$\Omega$ with an 
$\epsilon$ boundary layer. (a) No interior boundaries. (b) A pair of interior boundaries. Here $\x_k$ is the center of the $k^{\rm th}$ maximally extended circle which is a randomly selected point on the circumference of the $(k-1)^{\rm th} $ circle. The run terminates when a chosen point lands within the boundary layer. }
        \label{fig2}
    \end{figure}

    In Fig. \ref{fig2}, we illustrate schematically a single run of the WOS algorithm in domain contained zero or two interior boundaries. In both cases, the position of the particle is updated by randomly sampling a point on the surface of a maximally extended disk and then repeating the process until the particle enters a small neighborhood of the boundary. The total time to reach the boundary layer is calculated using an exact analytical formula for the survival probability of diffusion in a totally absorbing disk, as we now explain.
    
    Assume that at some time $t$,  we have $\X_t \in \Omega\backslash\calU_a$ and let 
    \begin{align}
        B_\rho(\X_t) =  \left\{\x \in \Omega:  \|\x - \X_t\| < \rho\right\},
    \end{align}
    where $\rho$ is taken to be sufficiently small so that $B_\rho(\X_t)\subset\Omega\backslash\calU_a$. The particle exits $B_\rho$ at time  $t + \calT_{\rho}$ where $\calT_{\rho}$ 
    is a random time with cumulative distribution function (CDF) $\mathcal{F}_\rho(\tau)$. 
    Since the particle diffuses without drift, $\X_{t + \calT_{\rho}}$ is uniformly distributed over $\partial B_\rho$.  Therefore, we only need to obtain $\calF_\rho(\tau)$
    to calculate the particle position and time after each WOS iteration. We can write 
    \begin{align}
        \calF_\rho(\tau) = \prob{\calT_\rho < t\middle|\X_0 = \bm 0} = 1 - \prob{\calT_\rho > t\middle|\X_0 = \bm 0} = 1 - Q_\rho(\bm 0, \tau),
    \end{align}
    where $Q_\rho(\x_0, \tau)$ is the survival probability of a Brownian particle diffusing in a disk $B_\rho(\bm 0)$ with a totally absorbing boundary $\partial B_\rho(\bm 0)$ 
    given that $\X_0=\x_0\in B_\rho(\bm 0)$.
    The probability density $p_\rho(\z, \tau|\x)$ of a Brownian particle diffusing in $B_\rho(\bm 0)$ satisfies the backward equation \cite{bressloff-book}
    \begin{subequations}
        \label{wos-backward}
        \begin{align}
            &\frac{\partial}{\partial t}p_\rho(\z, \tau|\x) = D\nabla^2_{\x} p_\rho(\z, \tau|\x), \ \x\in B_\rho(\bm 0),\\
            &p_\rho(\z, \tau|\x) = 0, \ \x\in\partial B_\rho(\bm 0)
        \end{align}
    \end{subequations}
    Integrating equation (\ref{wos-backward}) with respect to $\z$ over $B_\rho(\bm 0)$ yields the BVP 
    \begin{subequations}
        \label{survival-bvp}
        \begin{align}
            &\frac{\partial Q_\rho}{\partial \tau} = D\nabla^2Q_\rho, \ \x \in B_\rho(\bm 0),\\
            &Q_\rho(\x, \tau) = 0, \ \x \in \partial B_\rho(\bm 0), \\
            &Q_\rho(\x, 0) = 1, \ \x\in B_\rho(\bm 0) \label{survival-initial}
        \end{align}
    \end{subequations}
    If we rewrite the BVP (\ref{survival-bvp}) in terms of polar coordinates and set $Q_\rho(r, \tau) = \calR(r)\mathcal{E}(\tau)$, we obtain the eigenvalue problem
    \begin{align}
        &\frac{d^2\calR}{dr^2} + \frac{1}{r}\frac{d\calR}{dr} = \lambda\calR, \ r\in (0, \rho)\\
        &\calR(\rho) = 0, \ \calR(r) < \infty
    \end{align}
    The  eigenvalues and eigenfunctions are given by
    \begin{align}
        \calR_n(r) = J_0\left(\beta_n\frac{r}{\rho}\right), \ \lambda_n = -\left(\frac{\beta_n}{\rho}\right)^2, \ n \in \mathbb{N}\backslash\{0\},
    \end{align}
    where $J_m$ is the order $m$ Bessel function of the first kind and $\beta_n$ is the $n^{\text{th}}$ zero of $J_0$.
    Additionally, we have
    \begin{align}
        \frac{d\mathcal{E}_n}{d\tau} = -\left(\frac{\beta_n}{\rho}\right)^2\mathcal{E}_n.
    \end{align}
    Therefore,
    \begin{align}
        \mathcal{E}_n(\tau) = A_ne^{-(\beta_n/\rho)^2 \tau},
    \end{align}
    where $A_n$ are arbitrary coefficients determined by the initial condition (\ref{survival-initial}). The general solution to (\ref{survival-bvp}) is given by
    \begin{align}
       \label{survival-gen}
       Q_\rho(r, \tau) = \sum_{n=1}^\infty A_ne^{-(\beta_n/\rho)^2 \tau}J_0\left(\frac{\beta_n}{\rho} r\right).
    \end{align}
    Observe that
    \begin{align}
        \int_0^{\rho}xJ_0(ax)J_0(bx)dx = \frac{\rho\left[aJ_0(b\rho)J_1(a\rho) - bJ_0(a\rho)J_1(b\rho)\right]}{a^2 - b^2}
    \end{align}
    which implies that the eigenfunctions $\calR_n$ are orthogonal with respect to the weighted inner product
    \begin{align}
        \left<f, g\right> = \int_0^\rho xf(x)g(x)dx
    \end{align}
    where $f, g \in L^2([0, \rho])$. Therefore, substituting (\ref{survival-gen}) into (\ref{survival-initial}) and applying the weighted inner product yields
    \begin{align}
        A_n = \frac{1}{\left<J_0\left(\beta_n\frac{r}{\rho}\right), J_0\left(\beta_n\frac{r}{\rho}\right)\right>}\int_0^\rho rJ_0\left(\beta_n\frac{r}{\rho}\right)dr = \frac{2}{\beta_n J_1(\beta_n)}.
    \end{align}
    Therefore, we have 
    \begin{align}
        \label{unit-survival-prob}
        Q_\rho(\x, \tau) = \sum_{n=1}^{\infty}\frac{2}{\beta_n J_1(\beta_n)}e^{-(\beta_n/\rho)^2 \tau}J_0\left(\beta_n\frac{\|\x\|}{\rho}\right) = Q_0\left(\x/\rho, \tau/\rho^2\right).
    \end{align}
    
    We now introduce the boundary layers $\Omega_\delta$ and $\calU_{\delta, j}^\pm$ with width $\delta \ll 1$ 
    defined as 
    \begin{subequations}
        \begin{align}
            &\Omega_\delta = \left\{\x \in \Omega\backslash\calU_a: \text{dist}(\x, \partial\Omega\backslash\partial\calU_a) < \delta\right\},\\
            &\calU_{j, \delta}^+ = \left\{\x \in \Omega\backslash\calU_a: \text{dist}(\x, \partial\calU_j) < \delta\right\},\\
            &\calU_{j, \delta}^- = \left\{\x \in \calU_j: \text{dist}(\x, \partial\calU_j) < \delta\right\}.
        \end{align}
    \end{subequations}
    We can simulate the dynamics of the Brownian particle when $\X_t\notin\Omega_\delta,\calU_{j, \delta}^\pm$ by first computing 
    \begin{align}
        \label{wos-radius-outside} 
        \rho = \min\left\{\text{dist}(\X_t, \partial\Omega\backslash\partial\calU_a), \text{dist}(\X_t, \partial\calU_1), \ldots, \text{dist}(\X_t, \partial\calU_N)\right\}.
    \end{align}
    We then uniformly sample a number from the interval $\theta\in[0, 2\pi)$ and set 
    \begin{align}
        \label{wos-spatial-update}
        \X_{t + \calT_{\rho}} = \X_t + \Delta\X_t.
    \end{align}
    where
    \begin{align}
        \label{wos-dx}
        \Delta\X_t = \left(\rho\cos(\theta), \rho\sin(\theta)\right)^T.
    \end{align}
    Next, we sample the random variable $\calT_{\rho}$ by first sampling a standard uniform random variable $U$ and using the equations
    \begin{align}
        \label{exit-time-eqs}
        \calT_1 = \mathcal{F}_1^{-1}(U), \quad \calT_{\rho} = \rho^2\calT_1.
    \end{align}
    In practice, this is executed by storing an array of pre-computed values of $\mathcal{F}_1$ with a temporal resolution of $\delta t$ using a truncated version of equation (\ref{unit-survival-prob}).   
    After sampling $U$ using a standard random number generator, we apply a binary search algorithm to find the array element closest to $U$.
    After sampling $\calT_\rho$, we let $t \to t + \calT_{\rho}$ and repeat this process until the particle enters one of the boundary layers.  
    If $\X_t\in\calU_j\backslash\calU_{j, \delta}^-$, we let
    \begin{align}
        \label{wos-radius-inside}
        \rho = \text{dist}(\X_t, \partial\calU_j)
    \end{align}
    and compute $\calT_\rho$ and $\X_{t + \calT_\rho}$ using equations (\ref{wos-spatial-update}), (\ref{wos-dx}), and (\ref{exit-time-eqs}) as was done previously.

    \subsection{Skorokhod integrals and boundary local times}
    The dynamics of the particle in the boundary layers must be altered in order to simulate particle-boundary interactions and compute
    the boundary local times.  Once the particle enters a boundary layer, we set $\rho = 2\delta$ and then update the particle's position and time as 
    we did before but now there is a non-zero probability that the particle crosses a trap or domain boundary.  
    In the event that the particle crosses a boundary, the particle is projected onto the boundary along the normal vector.  
    Furthermore, we need a local time representation that generates accurate estimates using only the boundary-collision data generated from this simulated event.
    Such a representation can be obtained using the Skorokhod integral formulation of the local time \cite{elton}.

    Let $\{\Y_t\}_{t\geq 0}$ be a diffusion process on a bounded domain $D$, $\psi(\x, t)$ a real-valued non-negative function on $\partial D\times[0,\infty)$, and
    $\lambda$ a partition of the interval $[t_1, t_2]$ given by 
    \begin{align}
        \lambda: t_1=\tau_0<\tau_1<\cdots<\tau_n=t_2
    \end{align}
    Also, we define the size of the partition as
    \begin{align}
        d(\lambda) = \max_{0\leq j\leq n-1}\left[\tau_{j+1} - \tau_j\right]
    \end{align}
    Now consider the Riemann sum
    \begin{align}
        \label{reimann-sum}
        I(\lambda, \psi) = \sum_{j=0}^{n-1}\sqrt{\tau_{j+1} - \tau_j}\max_{s\in\lambda_j}\left[I_{\partial D}(\Y_s)\psi(\Y_{s}, s)\right]
    \end{align}
    If there exits a random variable $\mathcal{V}(\psi)$ such that  
    \begin{align}
        \lim_{\substack{d(\lambda)\to 0}}\expect{\left\|I(\lambda, \psi) - \mathcal{V}(\psi)\right\|^2} = 0,
    \end{align}
    then $\psi$ is a Skorokhod integrable function and the Skorokhod integral is denoted as
    \begin{align}
        \mathcal{V}(\psi) = \int_{t_1}^{t_2}\psi(\Y_\tau, \tau)\sqrt{d\tau}.
    \end{align}
    It can be shown that when $\psi(\Y_t, t) = \sqrt{\pi / 2}$, we have that
    \begin{align}
        \label{local-time-skorokhod}
        L^{\partial D}_t = \sqrt{\frac{\pi}{2}}\int_0^tI_{\partial D}(\Y_\tau)\sqrt{d\tau}.
    \end{align}
    where $L^{\partial D}_t$ is the boundary local time for the diffusion process $\{\Y_t\}_{t\geq 0}$ associated with the boundary $\partial D$.
    It follows from equations (\ref{local-time-skorokhod}) and (\ref{reimann-sum}) that the local times associated with the trap boundaries $\partial\calU_j$ can be expressed as 
    \begin{align}
        \label{local-time-alt}
        L^{(j)}_T &= D\sqrt{\frac{\pi}{2}}\int_0^TI_{\partial\calU_j}(\X_\tau)\sqrt{d\tau}\nonumber\\
        &\approx \delta D\sqrt{2\pi}\sum_{k=1}^{\mathcal{N}(T)}\sqrt{\calT_1^{(k)}}I_{\partial\calU_j}\left(\X_{t_{k-1} + \rho_{k}^2\calT_1^{(k)}}\right), \ j=1,\ldots,N.
    \end{align}
    where $\calT_1^{(k)}$ and $\rho_k$ are the random sample of $\calT_1$ and radius of the $k^{\text{th}}$ WOS iteration respectively 
    and $\mathcal{N}(T)$ is the number of WOS iterations performed before $t > T$.

   \begin{algorithm}[t!]
\caption{Simulation of SNOBM}\label{alg:gill}
\begin{algorithmic}[1]
\State Let $\calM$ be the region in which the particle is diffusing. That is, $\calM = \Omega\backslash\calU_a$ or $\calM = \calU_j$ for some $j=1,\ldots,N$. Set $\calM = \Omega\backslash\calU_a$, $t=0$, $\X_0 = \x\in\calM$, and $A_0^{(j)} = 0$ for all $j=1,\ldots,N$. Also, generate a random sample $\Ah_1,\ldots,\Ah_N$ of the occupation time thresholds
        $\Ah_j\sim\text{Exp}(\gamma)$, $j=1,\ldots,N$.
%\While{$t_k \le T$}
	\State Set $L^{(j)}_t = 0$ for all $j=1,\ldots,N$ and generate a sample $\Lh_1,\ldots,\Lh_N$ of the local time thresholds $\Lh_j\sim\text{Exp}(\kappa_j)$, $j=1,\ldots,N$. 
           \State  If $\X_t\in\Omega_\delta\cup\calU_{j, \delta}^{\pm}$, then set the WOS radius as $\rho = 2\delta$. Otherwise, compute $\rho$ using equation (\ref{wos-radius-outside}) 
        if $\calM = \Omega\backslash\calU_a$ or equation (\ref{wos-radius-inside}) if $\calM = \calU_j$ for some $j=1,\ldots,N$. 
        \State Generate a random sample of $\calT_\rho$ using equations (\ref{exit-time-eqs}) and compute $\X_{t + \calT_\rho}$ using equations (\ref{wos-spatial-update}) and (\ref{wos-dx}).
        If $\X_{t + \calT_\rho}\notin\calM$, project $\X_{t + \calT_\rho}$ to the nearest boundary point using the normal vector.
        \State  Calculate the occupation times using 
        \begin{align*}
            A_{t + \calT_\rho}^{(j)} = A_t^{(j)} + I_{\overline{\calU}_j}(\X_t)\calT_\rho, \ j = 1,\ldots,N.
        \end{align*}
        If $A_{t + \calT_\rho}^{(k)} \geq \Ah_k$ for some $k=1,\ldots,N$, then terminate the simulation and record 
        \begin{align*}
            \calT = t + \calT_\rho - A^{(k)}_{t + \calT_\rho} + \Ah_k
        \end{align*}
        as the FPT and $\calU_k$ as the absorbing trap.  Otherwise, proceed to step 6. 
        \State Calculate the local times using
        \begin{align*}
            L_{t + \calT_\rho}^{(j)} = L_t^{(j)} + I_{\partial\calU_j}\left(\X_{t + \calT_\rho}\right)\sqrt{\frac{\pi}{2}\calT_\rho}, \ j=1,\ldots,N.
        \end{align*}
        If $L_{t + \calT_\rho}^{(k)} \geq \Lh_k$ for some $k=1,\ldots,N$, proceed to step 7, otherwise proceed to step 8. 
        \State Generate a sample of $\mathcal{B}_k$.  If $B_k=0$ and $\calM = \Omega\backslash\calU_a$, then set $\calM = \calU_k$.
        If $B_k=1$ and $\calM = \calU_k$, then set $\calM = \Omega\backslash\calU_a$. Let $t\to t + \calT_\rho$ and return to step 2. 
        \State  Let $t\to t + \calT_\rho$ and return to step 3.
%\EndWhile
\end{algorithmic}
\end{algorithm}

 \subsection{Summary of the algorithm}    
    The basic steps of the stochastic simulation algorithm are summarized in Algorithm \ref{alg:gill}.
        The MFPT and splitting probabilities can then be estimated using the equations
    \begin{subequations}
        \label{mc-estimates}
        \begin{align}
            &\widehat{T} = \frac{1}{M}\sum_{i=1}^M\calT^{(i)}, \label{mc-mfpt}\\
            &\widehat{\pi}_k = \frac{1}{M}\sum_{i=1}^MI_{\overline{\calU}_k}\left(\X_{\calT^{(i)}}\right), \ k=1,\ldots,N \label{mc-sp}
        \end{align}
    \end{subequations}
    respectively where $M$ is the number of simulated trajectories and $\calT^{(i)}$ is the FPT of the $i^{\text{th}}$ trajectory.
   We remark that Algorithm \ref{alg:gill} could be modified to implement the local time method described in Ref. \cite{zhou} by setting $\rho=\Delta x$ when 
    $\Delta x < \text{dist}\left(\X_t, \partial\calU_j\right) < \epsilon$ and setting $\rho = 2\Delta x$ when 
    $\text{dist}\left(\X_t, \partial\calU_j\right) < \Delta x$ where $\Delta x = \epsilon / 3$.  The time between WOS iterations when the particle is in a boundary
    layer is estimated using the equation $\Delta t = \rho^2 / 4D$ which is based on the mean-squared-displacement formula for 2-D Brownian motion.
    The local time is computed using representation (\ref{local-time-main}) and the equation
    \begin{align}
        L^{(j)}_T &\approx \frac{D}{\epsilon}\int_0^TI_{\calU_{j,\delta}^\pm}(\X_t)dt 
            \approx \frac{1}{4D\epsilon}\sum_{k=2}^{\mathcal{N}(T)}\rho_k^2I_{\calU_{j,\delta}^\pm}(\X_{t_{k-1}}).
    \end{align}
    When the particle is outside the boundary layers, the trajectory is simulated as it was in algorithm (\ref{alg:gill}).

    \section{Numerical results}
    In this section, we test the accuracy and convergence properties of the algorithm developed in section 3.  The parameters that effect the accuracy 
    of the simulation are the boundary layer width, the number of realizations $M$, and the temporal resolution, $\delta t$, used when inverting the exit time CDF.
    Since the greatest source of error in the SNOBM simulations is the local time computations,  we first evaluate the accuracy of the local time 
    calculations independently of the rest of the algorithm. We proceed by solving for the MFPT to absorption of a Brownian particle in a disk with a partially reactive boundary. 
    Next, we test the efficacy of the full SNOBM simulations by comparing the results to equation (\ref{single-target-solution}).  Finally, we demonstrate the algorithm's ability 
    to solve narrow capture problems with semi-permeable partially absorbing traps by comparing it to the analytical solutions obtained using matched asymptotic
    and Green's function methods.
    All Monte Carlo algorithms were implemented as GPU kernels and executed on a NVIDIA Titan Volta GPU. 
    The GPU kernels were written using the {\em Numba Python} module which pre-compiles a restricted subset of python code into CUDA GPU kernels.
    
    \subsection{Boundary Local Time Calculations}
\begin{figure}[h]
        \centering
        \includegraphics[width=16cm]{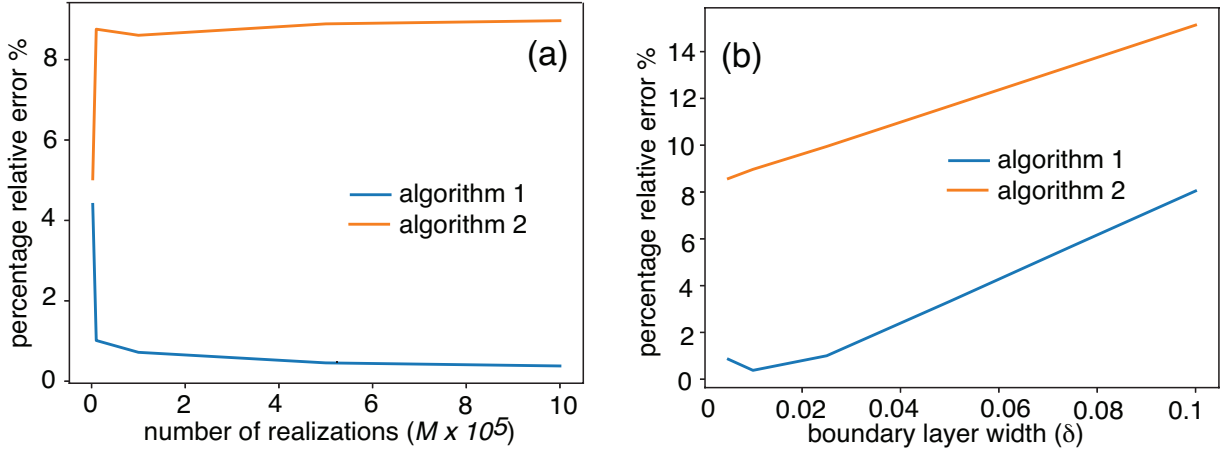} 
        \caption{Plots of the Monte Carlo estimated relative error with respect to the exact MFPT of equation (\ref{robin-mfpt}) as a function of (a) the number of Brownian motion realizations ($M$) for $\delta t = 0.01$, and (b) the boundary layer width $\delta$ for $M=10^6$, with $\kappa = 0.1$ and $D=1$. Algorithm (i) is our own algorithm based on the Skorokhod integral representation of the local time, see equation (\ref{local-time-alt}), whereas algorithm II is the one used in Ref. \cite{zhou}.}
        \label{local-time-num-sims}
    \end{figure}

    Consider a Brownian particle diffusing in the unit disk $S = \{\x\in\mathbb{R}^2: \|\x\|\leq 1\}$ where $\partial S$ acts as a partially reactive surface 
    with reactivity $\kappa$.  The Laplace transform of the probability density satisfies the following Robin BVP \cite{narrow-capture-robin}:
    \begin{subequations}
        \label{robin-bvp}
        \begin{align}
            &D\nabla^2_{\x} \widetilde{p}(\x, s|\y) - s\widetilde{p}(\x, s|\y) = -\delta(\x - \y), \ \|\x\| < 1,\\
            &-D\nabla_{\x} \widetilde{p}(\x, s|\y)\cdot\n = \kappa \widetilde{p}(\x, s|\y), \ \|\x\| = 1
        \end{align}
    \end{subequations}
   The BVP (\ref{robin-bvp}) can be solved exactly using circular symmetry and the Green's function of the modified Helmholtz equation.  Applying the relationship
    \begin{align}
        T(\y) = \left.\frac{\partial}{\partial s}\int_{\partial S}D\nabla\widetilde{p}(\x, s|\y)\cdot\n d\x\right|_{s=0}
    \end{align}
    yields the following equation for the MFPT
    \begin{align}
        \label{robin-mfpt}
        T(\y) = \frac{1 - \|\y\|^2}{4D} + \frac{1}{2\kappa}.
    \end{align}
    In figure \ref{local-time-num-sims}(a), we compare the accuracy of our local time algorithm and that of the algorithm developed by 
    Zhou {\em et al.} in \cite{zhou}.  We see that the relative error of our local time calculations monotonically decreases with the number of realizations used in 
    the Monte Carlo simulation and we are able to achieve a
    relative error less than $0.5\%$.  Counterintuitively, the error of the algorithm in Ref. \cite{zhou}increases for sufficiently large number of realizations and converges to a relative error just under $9\%$.
    This behavior indicates that the our algorithm is representing the boundary-particle interactions with much greater accuracy than the alternative method. Furthermore, we see from 
    figure (\ref{local-time-num-sims}(b)) that our algorithm performs better for all values of $\delta\in[0.005, 0.1]$.

    \subsection{Full SNOBM Simulations}
\begin{figure}[h!]
        \centering
        \includegraphics[width=16cm]{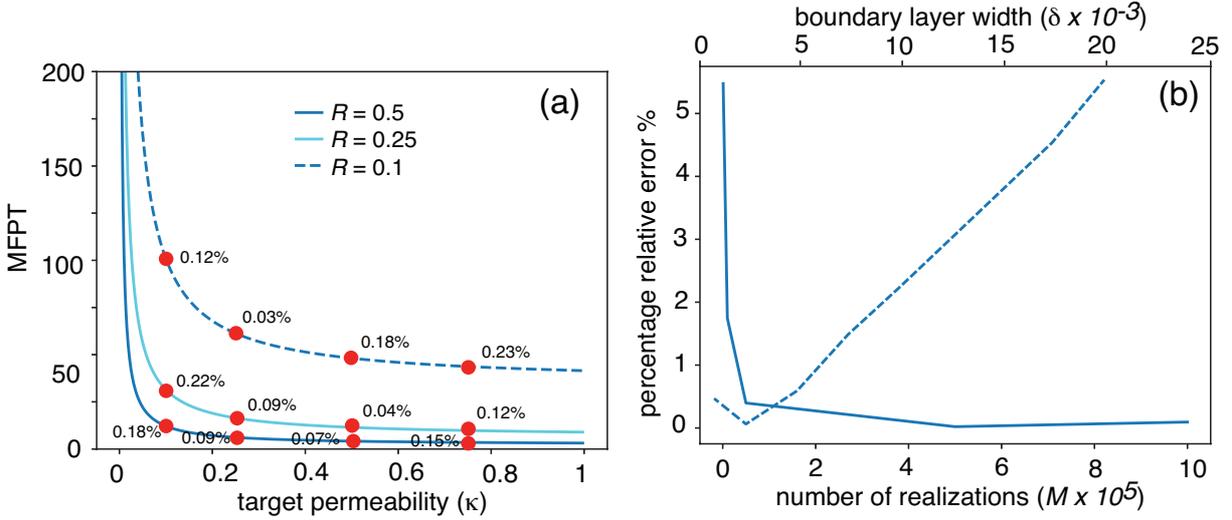} 
        \caption{(a) Plots of MFPT given by equation (\ref{single-target-solution}) as a function of the permeability $\kappa$ for various  values of the radius $R$ with $\gamma = 1.0$, $D_0=D_1 = 1.0$, $\alpha = 0.25$, $\delta = 2.5\times 10^{-3}$, $M=10^7$, and $\x_0 = (0.75, 0)^T$. Additionally, estimates of the MFPT using (\ref{mc-estimates}) for $\kappa = 0.1, 0.25, 0.75$ are provided with relative error calculations. (b) Plots of the relative error of (\ref{mc-estimates}) with respect to equation (\ref{single-target-solution}) as a function of the number of realizations $M$ with $\delta=2.5\times 10^{-3}$ (solid curve) and the boundary layer width $\delta$ with $M=10^6$ (dashed curve). The other parameter values are $\kappa=0.25$, $\gamma=1$, $\alpha=0.25$,
                 $D_0=D_1=1$, $\X_0 = (0.75, 0)^T$, and $R=0.25$.}
        \label{single-target-kappa}
    \end{figure}

    We now analyze the accuracy of the full SNOBM simulations by comparing the Monte Carlo approximation of the MFPT (\ref{mc-estimates}) with 
    the exact MFPT (\ref{single-target-solution}) for a Brownian particle diffusing in the unit disk with a single circular trap centered at the origin. 
    In Fig. \ref{single-target-kappa}(a), we plot the exact MFPT as a function of the boundary reactivity $\kappa$.  We also mark the Monte Carlo estimates
    of the MFPT for a subset of the $\kappa$-values. For $\delta=2.5\times 10^{-3}$, $M=10^{-7}$, and $\delta t = 0.01$,
    the relative error of the Monte Carlo estimates is less than $0.25\%$ even when $R \ll 1$. In Fig.\ref{single-target-num-sims}(b)
    we plot the relative error of the Monte Carlo estimates against the parameters $M$ and $\delta$. We see that the simulated MFPTs are within $0.5\%$ of the true 
    values for $M\geq 5\times 10^4$ when $\delta=2.5\times 10^{-3}$.  Additionally, the Monte Carlo estimates achieve relative errors of less than $1\%$ when $\delta\leq 0.005$.  
    Note that there is a slight increase in error when we decrease $\delta$ from $2.5\times 10^{-3}$ to $10^{-3}$.  This is likely due to the fact that the variance of the 
    Monte Carlo estimates increases as $\delta$ decreases and the fact that the probability of the simulated trajectory reaching the boundary layer gets too small.

    \subsection{A 2D Narrow Capture Problem}
\begin{figure}[t!]
        \centering
        \includegraphics[width=16cm]{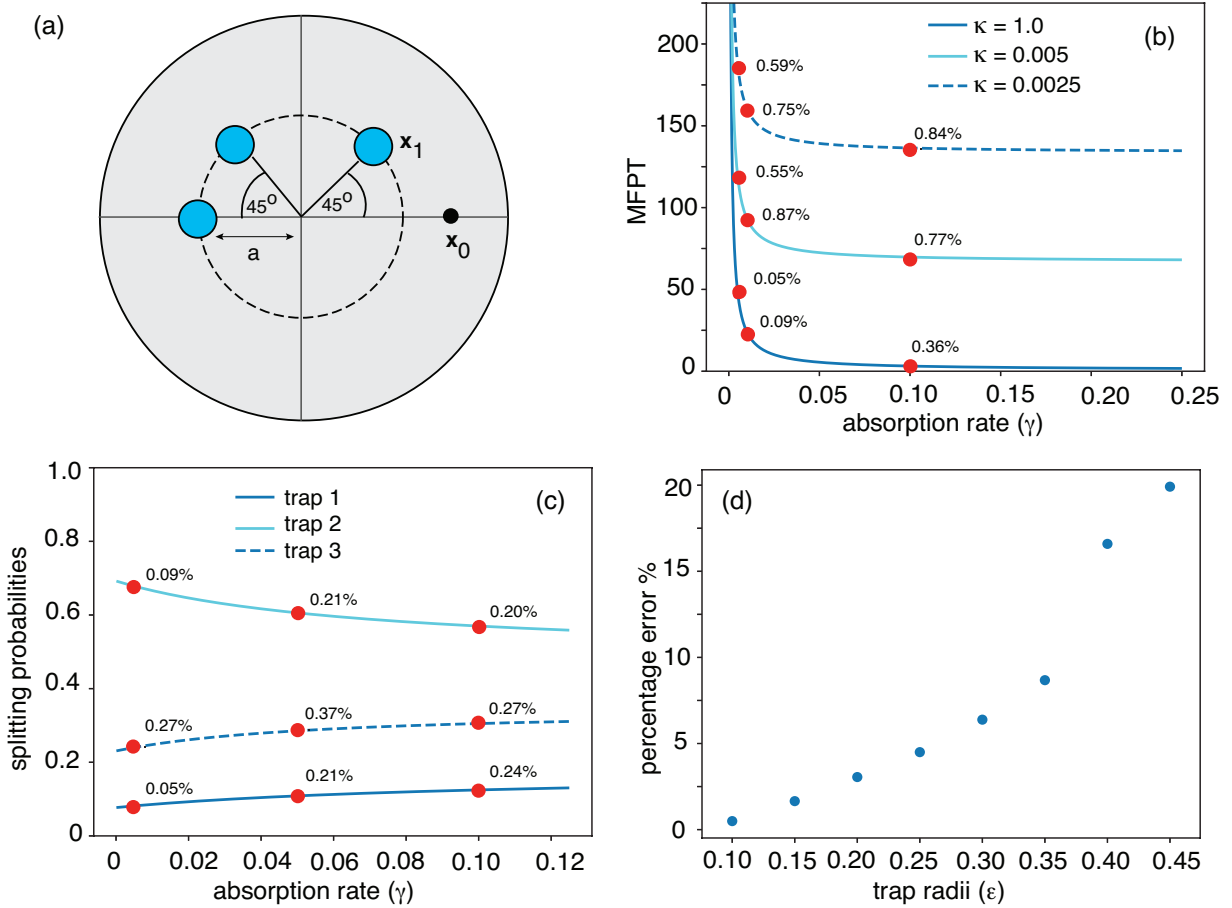} 
        \caption{Unit disk containing three small traps. (a) Basic trap configuration with centers at $\x_j$, $j=1,2,3$. The particle starts at the point $\x_0$. (b) Plots of the MFPT given by equation (\ref{mfpt-asymptotic}) for various $\kappa'$ values where $\kappa'_j = \kappa'$ for all $j=1,2,3$. The other parameter values are $D_j=D=1$, $\epsilon=0.075$, $\rho_j=1$, $\alpha_1=0.75$, $\alpha_2=0.25$, and $\alpha_3=0.5$. 
                 Additionally, Monte Carlo estimates of the MFPT using equation (\ref{mc-mfpt}) are provided with relative error calculations for $\gamma'=0.005,0.01,0.1$, $\delta=2.5\times 10^{-3}$ and $M = 10^7$. (c) Corresponding plots of the splitting probabilities given by equation (\ref{sp-asymptotic}) with $\kappa' = 0.1$. 
                 Monte Carlo estimates of the MFPT using equation (\ref{mc-sp}) are provided with relative error calculations for $\gamma'=0.005,0.05,0.1$
                 with $\delta=2.5\times 10^{-3}$ and $M = 10^7$. (d) Corresponding plot of the percent error of equation (\ref{mfpt-asymptotic}) with respect to equation (\ref{mc-mfpt}) as a function of $\epsilon$ for $\kappa' = 0.1$, $\gamma'=0.01$, and $M=10^6$. }
        \label{fig5}
    \end{figure}

    As our final example, we apply the SNOBM algorithm to the unit disk $\Omega$ containing three traps whose centers are located at $\x_1 = \left(\sqrt{2} / 4, \sqrt{2} / 4\right)^T$, $\x_2 = \left(-\sqrt{2} / 4, \sqrt{2} / 4\right)^T$,
    and $\x_3 = \left(0, -1/2\right)^T$, see Fig. \ref{fig5}(a). The MFPT and splitting probability 
    BVPs are too complex to solve exactly, so we use asymptotic analysis to obtain approximate solutions, under the assumptions that 
    the area of each trap is $\calO(\epsilon^2)$, $\kappa_j = \calO(\epsilon^{-1})$, and $\gamma = \calO(\epsilon^{-2})$ where $0<\epsilon\ll 1$. The details of the asymptotic analysis are presented in the appendix.
    
        In Fig. \ref{fig5}(b,c), we compare the MFPTs and splitting probabilities obtained 
    using the Monte Carlo estimates (\ref{mc-estimates}) and the asymptotic approximations (\ref{sp-asymptotic}) and (\ref{mfpt-asymptotic}). We see that for 
    $\epsilon = 0.075$, the asymptotic results for both the MFPT and splitting probabilities deviate from the Monte Carlo simulations by less than $1\%$. Furthermore, Fig. 
    \ref{fig5}(d) indicates that we can achieve errors under $3\%$ for all $\epsilon \leq 0.2$. These results show that our Monte Carlo algorithm provides a useful numerical 
    approach to validating narrow capture asymptotic solutions and exploring parameter regimes in which the asymptotic solutions break down.

    \section{Conclusion}
    In this paper, we developed a Monte Carlo algorithm for solving SNOBM and computing the MFPT and splitting probabilities for reaction-diffusion 
    processes involving semi-permeable partially absorbing traps.  The numerical methods were shown to have high accuracy when compared 
    to solutions obtained using BVP methods and matched asymptotic analysis.  Furthermore, the method we developed to compute boundary local times 
    significantly out performs previous methods for multi-dimensional domains. This indicates that our algorithm has applications beyond simulating SNOBM.  
    For example, single particle reaction-diffusion processes with generalized partially reactive surfaces involving non-Markovian absorption processes \cite{bressloff-bf} 
    could be simulated by adapting the methods presented here. Additionally, the solutions to elliptic and parabolic Neumann boundary value problems can be represented as 
    Brownian functionals involving boundary local time integrals \cite{zhou,sde-pde}.  
    Therefore, our local time algorithm could be integrated into a Monte Carlo PDE solver.
    Traditional numerical methods for solving PDEs require
    one to partition the domain into a grid and estimate the solution at each grid point even if the solution only needs to be evaluated at a small subset of points.  
    Also, the time complexity of standard numerical solvers scale exponentially
    with the dimension of the PDE making them inefficient for high dimensional BVPs.  Monte Carlo based numerical methods solve both of these limitations by leveraging 
    the stochastic representation of the solution and the parallelizability of Monte Carlo simulations. Finally, we note that one can extend our methods to 3-D reaction-diffusion
    processes by making two modifications.  First, both polar and azimuthal angles needed to be sampled for each WOS iteration to compute the particle position.  
    That is, equation (\ref{wos-spatial-update}) becomes
    \begin{align}
        \Delta\X = \left(\rho\sin(\phi)\cos(\theta), \rho\sin(\phi)\sin(\theta), \rho\cos(\phi)\right)^T
    \end{align}
    where $\phi$ and $\theta$ are uniformly sampled from the intervals $[0, \pi]$ and $[0, 2\pi)$ respectively.
    Second, 
    the solution to the survival probability BVP (\ref{survival-bvp}) becomes 
    \begin{align}
        Q_\rho(\x, t) = \sum_{n=1}^\infty\frac{2(-1)^{n+1}}{n\pi r}\sin\left(n\pi\frac{r}{\rho}\right)e^{-\left(n\pi / \rho\right)^2t}.
    \end{align}

    \appendix
    
    \section{Matched asymptotic analysis}
    Consider the Brownian particle described in section 2. 
    We define the narrow capture problem for multiple semi-permeable partially absorbing traps as follows: Each trap is assumed to have an area 
    $|\calU_j|\sim \epsilon^2 |\Omega|$ with $\calU_j\to\x_j\in \Omega$ uniformly as $\epsilon \to 0$. 
    The traps are also assumed to be well separated such that $\text{dist}(\x_j ,\partial \Omega)=\calO(1)$ and $\|\x_j-\x_i\|=\calO(1)$ for all $i,j =1,\ldots,N$ and $j\neq i$.  
    For concreteness, we will take each trap to be a disk of radius $r_j=\epsilon \rho_j$. Thus, $\calU_j=\{\x \in \Omega: \|\x-\x_j\|\leq \epsilon \rho_j\}$.  
    In order to maintain effective absorption in the limit $\epsilon\to 0$, we take $\gamma = \gamma'/\epsilon^2$ and $\kappa_j = \kappa'_j/\epsilon$ where $\gamma'$ and $\kappa'_j$ are $\calO(1)$
    constants. In this section, we solve the BVPs (\ref{sp-bvp}) and (\ref{mfpt-bvp}) for the narrow capture problem using matched asymptotic analysis and Green's function methods along
    the lines of \cite{narrow-capture-reset}.
    
    \subsection{Splitting Probabilities}
    Consider the asymptotic expansions
    \begin{align}
        \phi_k(\x) \sim \phi_k^{(0)}(\x) + \epsilon\phi_k^{(1)}(\x) + \calO(\epsilon^2), \ \Phi_k(\y) \sim \Phi_k^{(0)}(\y) + \epsilon\Phi_k^{(1)}(\y) + \calO(\epsilon^2)
    \end{align}
    where $\Phi_k(\y) = \pi_k(\x_j + \epsilon\y)$ is the inner solution with the stretch coordinate $\y = \epsilon^{-1}\left(\x - \x_j\right)$ and
    $\phi_k$ is the outer solution which satsifes 
    \begin{subequations}
        \label{sp-outer-bvp}
        \begin{align}
            &\nabla^2\phi_k^{(n)}(\x) = 0, \ \x\in\Omega\backslash\{\x_1,\ldots,\x_N\},\\
            &\nabla\phi_k^{(n)}(\x)\cdot\n_0 = 0, \ \x\in\partial\Omega,\\
            &\phi_k \sim \Phi_k, \ \x\to\x_j. \label{sp-match}
        \end{align}
    \end{subequations}
      Since the trap regions and the boundary are well separated, we can assume $\Phi_k^{(n)}(\y)$ is circularly symmetric.  Therefore, we have that 
    \begin{subequations}
        \label{sp-inner-bvp}
        \begin{align}
            &\frac{\partial^2\Phi_k^{(n)}}{\partial r^2} + \frac{1}{r}\frac{\partial\Phi_k^{(n)}}{\partial r} = 0, \ r > \rho_j, \label{sp-inner-plus} \\
            &\frac{\partial^2\Phi_k^{(n)}}{\partial r^2} + \frac{1}{r}\frac{\partial\Phi_k^{(n)}}{\partial r} - \frac{\gamma'}{D_j}\Phi_k^{(n)} = -\frac{\gamma'}{D_j}\delta_{k,j}, \ r < \rho_j,
                \label{sp-inner-minus}\\
            &\alpha_jD\left.\frac{\partial\Phi_k^{(n)}}{\partial r}\right|_{r=\rho_j^+} = (1 - \alpha_j)D_j\left.\frac{\partial\Phi_k^{(n)}}{\partial r}\right|_{r=\rho_j^-}
            = \kappa'_j\alpha_j(1 - \alpha_j)\left[\Phi_k^{(n)}(\rho^+_j) - \Phi_k^{(n)}(\rho^-_j)\right] \label{sp-inner-bc}
        \end{align}
    \end{subequations}
    It follows that the general solution to the $\calO(1)$ equation is given by
    \begin{gather}
        \label{inner-gen-sol}
        \Phi_k^{(0)}(r) =
        \begin{cases}
            C_0 + C_1\ln\left(r / \rho_j\right), \ r > \rho_j\\
            \delta_{k, j} + C_2I_0\left(\sqrt{\gamma' / D_j}r\right), r < \rho_j
        \end{cases}
    \end{gather}
    where $I_0$ is the modified Bessel Function of order zero and $C_0$, $C_1$, and $C_2$ are arbitrary constants. 
    Substituting (\ref{inner-gen-sol}) into (\ref{sp-inner-bc}) and solving for $C_0$, $C_1$, and $C_2$ yields
    \begin{gather}
        \label{sp-inner-solution}
        \Phi_k^{(0)}(\x) = 
        \begin{cases}
            \delta_{k, j} + \calA_{k, j}(\nu) + \nu\calA_{k, j}(\nu)\left[\calC_j + \ln\left(\rho_j^{-1}\|\x - \x_j\|\right)\right], \ \|\x - \x_j\| > \epsilon\rho_j,\\
            \delta_{k, j} + \nu\calA_{k, j}(\nu)\calB_jI_0\left(\sqrt{\gamma'/D_j}\|\x - \x_j\|/\epsilon\right), \ \|\x - \x_j\| < \epsilon\rho_j
        \end{cases}
    \end{gather}
    where $\nu = -1/\ln(\epsilon)$, $\calA_{k, j}(\nu)$ are coefficients determined by the match condition (\ref{sp-match}), and
    \begin{subequations}
        \begin{align}
            &\calC_j = \frac{D}{\rho_j(1 - \alpha_j)}\left[\frac{1}{\kappa'_j}\right.
                \left. + \frac{\alpha_j}{\sqrt{\gamma' D_j}}\frac{I_0\left(\sqrt{\gamma' / D_j}\rho_j\right)}{I_1\left(\sqrt{\gamma' / D_j}\rho_j\right)}\right], \\
            &\calB_j = \frac{\alpha D}{\rho_j(1 - \alpha_j)\sqrt{\gamma' D_j}I_1\left(\sqrt{\gamma'/D_j}\rho_j\right)}
        \end{align}
    \end{subequations}
    Equation (\ref{sp-inner-solution}) suggests the follwing ansatz for the outer solution
    \begin{align}
        \label{sp-asymptotic}
        \phi_k^{(0)}(\x) = \chi_k - 2\pi\nu\sum_{j=1}^N\calA_{k, j}(\nu)G(\x, \x_j)
    \end{align}
    where $\chi_k$ is an arbitrary constant and $G(\x, \x_j)$ is the the 2D Neumann Green's function which satisfies
    \begin{subequations}
        \label{green-bvp}
        \begin{align}
            &\nabla^2_{\x}G(\x, \y) = \frac{1}{|\Omega|} - \delta(\x - \y), \ \x, \y \in \Omega,\\
            &\nabla_{\x} G(\x, \y)\cdot\n_0 = 0, \ \x \in \partial\Omega, \ \y \in \Omega\\
            &\int_\Omega G(\x, \y)d\x = 0, \label{green-const} \\
            &G(\x, \y) = -\frac{1}{2\pi}\log\|\x - \y\| + R(\x, \y).
        \end{align}
    \end{subequations}
    Here, $R(\x, \y)$ is the regular part of the Green's function.  The solution to (\ref{green-bvp}) is uniquely defined and an exact formula can be obtained when 
    $\Omega$ is a disc with radius $R$ (appendix C). Applying the match condition (\ref{sp-match}) near the $i^{\text{th}}$ trap yields
    \begin{align}
        \chi_k - 2\pi\nu\calA_{k, i}(\nu)R(\x_i, \x_i) - 2\pi\nu\sum_{j\neq i}\calA_{k, j}(\nu)G(\x_i, \x_j) = \delta_{k, i} + \calA_{k, i}(\nu) + \nu\calA_{k, i}(\nu)\calC_i.
    \end{align}
    Therefore, we have that 
    \begin{align}
        \label{sp-coefs}
        \calA_{k, i}(\nu) = \sum_{j=1}^N\calH^{-1}_{ij}\bm{\chi}_j
    \end{align}
    where 
    \begin{align}
        \calH = \bm{I} + \nu\bm D + 2\pi\nu\calG, 
    \end{align}
    $\bm{I}$ is the $N\times N$ identity matrix,
    \begin{gather}
        \calG_{ij} = 
        \begin{cases}
            G(\x_i, \x_j), \ i\neq j,\\
            R(\x_i, \x_i), \ i = j,
        \end{cases}
        \quad
        \bm{\chi}_i = 
        \begin{cases}
            \chi_k, \ i\neq k,\\
            \chi_k - 1, \ i = k,
        \end{cases}
    \end{gather}
    and $\bm{D} = \text{diag}(\calC_1,\ldots,\calC_N)$. Note that
    \begin{align}
        \nabla^2\phi_k^{(0)} = 2\pi\nu\sum_{j=1}^N\calA_{k, j}(\nu)\delta(\x - \x_j) - \frac{2\pi\nu}{|\Omega|}\sum_{j=1}^N\calA_{k, j}(\nu).
    \end{align}
    Equation (\ref{sp-outer-bvp}) requires that 
    \begin{align}
        \sum_{j=1}^N\calA_{k, j}(\nu) = 0.
    \end{align}
    Therefore, summing equation (\ref{sp-coefs}) with respect to $i$ and solving for $\chi_k$ gives
    \begin{align}
        \chi_k = \frac{\sum_{i=1}^N\calH^{-1}_{ik}}{\sum_{j=1}^{N}\sum_{i=1}^N\calH^{-1}_{ij}}
    \end{align}

    \subsection{Mean First Passage Time}
    Consider the asymptotic expansions
    \begin{align}
        z(\x) \sim z^{(0)}(\x) + \epsilon z^{(1)}(\x) + \calO(\epsilon^2), \ \Z(\y) \sim \Z^{(0)}(\y) + \epsilon\Z^{(1)}(\y) + \calO(\epsilon^2)
    \end{align}
    where $\Z(\y) = T(\x_j + \epsilon\y)$  and $z(\x)$ are the inner and outer solutions of (\ref{mfpt-bvp}) respectively.  The outer solution satisfies
    \begin{subequations}
        \label{mfpt-outer-bvp}
        \begin{align}
            &\nabla^2z^{(n)}(\x) = -\frac{\delta_{n, 0}}{D}, \ \x\in\Omega\backslash\{\x_1,\ldots,\x_N\},\\
            &\nabla z^{(n)}(\x)\cdot\n_0 = 0, \ \x\in\partial\Omega,\\
            &z \sim \Z, \ \x\to\x_j \label{mfpt-match}
        \end{align}
    \end{subequations}
    and the inner solution satisfies
    \begin{subequations}
        \label{mfpt-inner-bvp}
        \begin{align}
            &\frac{\partial^2\Z^{(n)}}{\partial r^2} + \frac{1}{r}\frac{\partial\Z^{(n)}}{\partial r} = -\frac{\delta_{n, 2}}{D}, \ r > \rho_j, \label{mfpt-inner-plus} \\
            &\frac{\partial^2\Z^{(n)}}{\partial r^2} + \frac{1}{r}\frac{\partial\Z^{(n)}}{\partial r} - \gamma'\Z^{(n)} = -\frac{\delta_{n, 2}}{D_j}, \ r < \rho_j,
                \label{mfpt-inner-minus}\\
            &\alpha_jD\left.\frac{\partial\Z^{(n)}}{\partial r}\right|_{r=\rho_j^+} = (1 - \alpha_j)D_j\left.\frac{\partial\Z^{(n)}}{\partial r}\right|_{r=\rho_j^-}
            = \kappa_j'\alpha_j(1 - \alpha_j)\left[\Z^{(n)}(\rho^+_j) - \Z^{(n)}(\rho^-_j)\right] \label{mfpt-inner-bc}
        \end{align}
    \end{subequations}
    Therefore, we have that
    \begin{gather}
        \label{mfpt-inner-gen-sol}
        \Z^{(0)}(r) =
        \begin{cases}
            C_0 + C_1\ln(r / \rho_j), \ r > \rho_j\\
            C_2I_0\left(\sqrt{\gamma' / D_j}r\right), r < \rho_j
        \end{cases}
    \end{gather}
    Substituting (\ref{mfpt-inner-gen-sol}) into (\ref{mfpt-inner-bc}) and solving for $C_0$, $C_1$, and $C_2$ yields
    \begin{gather}
        \Z^{(0)}(\x) = 
        \begin{cases}
            \calF_j(\nu) + \nu\calF_j(\nu)\left[\calC_j + \ln\left(\rho_j^{-1}\|\x - \x_j\|\right)\right], \ \|\x - \x_j\| > \epsilon\rho_j,\\
            \nu\calF_j(\nu)\calB_jI_0\left(\sqrt{\gamma'/D_j}\|\x - \x_j\|/\epsilon\right), \ \|\x - \x_j\| < \epsilon\rho_j
        \end{cases}
    \end{gather}
    where $\calF_j(\nu)$ are coefficients determined by the match condition (\ref{mfpt-match}). The outer solution takes the form
    \begin{align}
        \label{mfpt-asymptotic}
        z^{(0)}(\x) = \xi - 2\pi\nu\sum_{j=1}^N\calF_j(\nu)G(\x, \x_j)
    \end{align}
    where $\xi$ is an arbitrary constant.
    Applying the match condition (\ref{mfpt-match}) near the $i^{\text{th}}$ trap yields
    \begin{align}
        \xi - 2\pi\nu\calF_i(\nu)R(\x_i, \x_i) - 2\pi\nu\sum_{j\neq i}\calF_j(\nu)G(\x_i, \x_j) = \calF_i(\nu) + \nu\calF_i(\nu)\calC_i.
    \end{align}
    Therefore, we have that 
    \begin{align}
        \label{mfpt-coefs}
        \calF_i(\nu) = \xi\sum_{j=1}^N\calH^{-1}_{ij}
    \end{align}
    Observe that
    \begin{align}
        \nabla^2z^{(0)} = 2\pi\nu\sum_{j=1}^N\calF_j(\nu)\delta(\x - \x_j) - \frac{2\pi\nu}{|\Omega|}\sum_{j=1}^N\calF_j(\nu).
    \end{align}
    Equation (\ref{mfpt-outer-bvp}) requires that 
    \begin{align}
        \sum_{j=1}^N\calF_j(\nu) = \frac{|\Omega|}{2\pi\nu D} 
    \end{align}
    Therefore, summing equation (\ref{sp-coefs}) with respect to $i$ and solving for $\xi$ gives
    \begin{align}
        \xi = \frac{|\Omega|}{2\pi\nu D}\left[\sum_{i=1}^N\sum_{j=1}^N\calH_{ij}^{-1}\right]^{-1}.
    \end{align}
    
    \subsection{The Neumann Green's Function in a Disk}
    Here, we calculate the Neumann Green's function in a disk with an arbitrary radius.  An analogous calculation for the 3D case can be found here \cite{ward-gf}.

    Let $\Omega$ be a disc with radius $R$. The solution to (\ref{green-bvp}) can be decomposed as 
    \begin{align}
        \label{green-decomp}
        G(\x, \y) = \Lambda - \frac{1}{2\pi}\log\|\x - \y\| + \frac{1}{4\pi}\left[\|\x\|^2 + \|\y\|^2\right] + \frac{1}{2\pi}\mu(\x, \y)
    \end{align}
    where $\Lambda$ is a constant determined by condition (\ref{green-const}) and
    \begin{subequations}
        \label{mu-bvp}
        \begin{align}
            &\nabla^2_{\x}\mu(\x, \y) = 0, \ \x, \y \in \Omega,\\
            \label{mu-bc}
            &D\nabla_{\x}\mu(\x, \y)\cdot\n_0 = \left[\frac{\x - \y}{\|\x - \y\|^2} - \frac{\pi}{|\Omega|}\x\right]\cdot\n_0 \equiv \Gamma(\x, \y) , \ \x\in\partial\Omega, \ \y \in \Omega.
        \end{align}
    \end{subequations}
    With out loss of generality, we can take $\y$ to be on the $x$-axis.  Writing (\ref{mu-bvp}) in terms of polar coordinates and assuming that $\mu(r, \theta) = \calR(r)\Theta(\theta)$ 
    yields the eigenvalue problem
    \begin{subequations}
        \label{eigen}
        \begin{align}
            &\frac{d^2 \Theta}{d\theta^2} = \lambda\Theta, \ \theta \in (-\pi, \pi), \\
            &\Theta(\pi) - \Theta(-\pi) = \left.\frac{d\Theta}{d\theta}\right|_{\theta=\pi} - \left.\frac{d\Theta}{d\theta}\right|_{\theta=-\pi} = 0
        \end{align}
    \end{subequations}
    and the differential equation
    \begin{align}
        \frac{d^2\calR}{dr^2} + \frac{1}{r}\frac{d\calR}{dr} + \lambda\calR = 0, \ \calR(r) < \infty, \ r \in [0, R]
    \end{align}
    Therefore, we have that
    \begin{align}
        \Theta_n(\theta) = \cos(n\theta), \ \lambda_n = -n^2, \ n = 0,1,2,\ldots.
    \end{align}
    and
    \begin{align}
        \calR_n(r) = A_nr^n
    \end{align}
    where $A_n$ are coefficients determined by the boundary condition (\ref{mu-bc}). The general solution to (\ref{mu-bvp}) can be written as 
    \begin{align}
        \label{mu-gen}
        \mu(r, \theta) = A_0 + \sum_{n=1}^{\infty}A_nr^n\cos(n\theta).
    \end{align}
    Note that $A_0$ is arbitrary so we can set $A_0=0$. Substituting (\ref{mu-gen}) into (\ref{mu-bc}) and using the orthogonality of the eigenfunctions $\Theta_n$ gives 
    \begin{align}
        A_n = \frac{1}{\pi nR^{n-1}}\int_{-\pi}^\pi\frac{\left(R + \|\y\|\cos\theta\right)\cos(n\theta)}{R^2 + \|\y\|^2 - 2\|\y\|R\cos\theta}d\theta = \frac{1}{n}\left(\frac{\|\y\|}{R^2}\right)^n
    \end{align}
    It follows that
    \begin{align}
        \mu(\x, \y) = \sum_{n=1}^\infty\frac{\cos(n\theta)}{n}\left(\frac{\|\y\|\|\x\|}{R^2}\right)^n.
    \end{align}
      Using the fact that
    \begin{align}
        \sum_{n=1}^\infty\frac{z^n}{n} = \log\left(\frac{1}{1 - z}\right), \ |z| < 1,
    \end{align}
    and setting
    \begin{align}
        \alpha = \frac{\|\x\|\|\y\|}{R^2}e^{i\theta},
    \end{align}
    we can write
    \begin{align}
        \label{mu-sol}
        \mu(\x, \y) &= \frac{1}{2}\sum_{n=1}^\infty\frac{\alpha^n}{n} + \frac{1}{2}\sum_{n=1}^\infty\frac{(\alpha^*)^n}{n}\nonumber\\
        &= \frac{1}{2}\log\left(\frac{1}{|\alpha|^2 - (\alpha + \alpha^*) + 1}\right)\nonumber\\
        &= -\log\sqrt{1 - \frac{2(\x\cdot\y)}{R^2} + \frac{\|\x\|^2\|\y\|^2}{R^4}}
    \end{align}
    Substituting (\ref{mu-sol}) into (\ref{green-decomp}), setting $\y=\bm{0}$, and integrating over $\Omega$ with respect to $\x$ yields
    \begin{align}
        0 = \int_\Omega G(\x, \bm{0})d\x = \frac{R^2}{2}\left[\frac{3}{4} - \log(R)\right] + \pi R^2\Lambda
    \end{align}
    Thus,
    \begin{align}
        \Lambda = \frac{1}{2\pi}\left[\log(R) - \frac{3}{4}\right].
    \end{align}


\begin{thebibliography}{10}
\expandafter\ifx\csname url\endcsname\relax
  \def\url#1{\texttt{#1}}\fi
\expandafter\ifx\csname urlprefix\endcsname\relax\def\urlprefix{URL }\fi
\expandafter\ifx\csname href\endcsname\relax
  \def\href#1#2{#2} \def\path#1{#1}\fi

\bibitem{zhou}
Y.~Zhou, W.~Cai, E.~Hsu, Computation of the boundary local time of reflecting
  brownian motion and the probabilistic representation of the neumann problem,
  Communications in Mathematical Sciences 15 (2017) 237--259.

\bibitem{bio-app-1}
R.~Phillips, J.~Kondev, J.~Theriot, H.~G. Garcia, N.~Orme, {Physical Biology of
  the Cell}, Garland Science, New York, 2012.

\bibitem{bio-app-2}
B.~Alberts, A.~Johnson, J.~Lewis, D.~Morgan, M.~Raff, K.~Roberts, P.~Walter,
  {Molecular biology of the cell, 6th ed.}, Garland Science, New York, 2015.

\bibitem{bressloff-book}
P.~C. Bressloff, {Stochastic Processes in Cell Biology, 2nd ed.}, Springer,
  Switzerland, 2021.

\bibitem{bio-app-3}
V.~Nikonenko, N.~Pismenskaya, Ion and molecule transport in membrane systems
  (special issue), Int. J. Mol. Sci. 22 (2021) 3556.

\bibitem{Brink85}
P.~R. Brink, S.~V. Ramanan, A model for the diffusion of fluorescent probes in
  the septate giant axon of earthworm: axoplasmic diffusion and junctional
  membrane permeability, Biophys. J. 48 (1985) 299--309.

\bibitem{Connors04}
B.~W. Connors, M.~A. Long, Electrical synapses in the mammalian brain, Ann.
  Rev. Neurosci. 27 (2004) 393--418.

\bibitem{Bressloff16}
P.~C. Bressloff, Diffusion in cells with stochastically-gated gap junctions,
  SIAM J. Appl. Math. 76 (2016) 1658--1682.

\bibitem{Grossel98}
P.~Grossel, F.~Depasse, Alternating heat diffusion in thermophysical depth
  profiles: multilayer and continuous descriptions, J. Phys. D: Appl. Phys. 31
  (1998) 216.

\bibitem{deMonte00}
F.~de~Monte, Transient heat conduction in one-dimensional composites lab. a
  natural analytic approach., Int. J. Heat Mass Transf. 43 (2000) 3607--3619.

\bibitem{Lu05}
X.~Lu, P.~Tervola, Transient heat conduction in the composites lab-analytical
  method, J. Phys. A: Math. Gen. 38 (2005) 81.

\bibitem{Tanner78}
J.~E. Tanner, Transient diffusion in a system partitioned by permeable
  barriers: application to nmr measurements with a pulsed field gradient., J.
  Chem. Phys. 69 (1978) 1748.

\bibitem{Callaghan92}
P.~T. Callaghan, A.~Coy, T.~P.~J. Halpin, D.~MacGowan, K.~J. Packer, F.~O.
  Zelaya, Diffusion in porous systems and the influence of pore morphology in
  pulsed gradient spin-echo nuclear magnetic resonance studies, J. Chem. Phys.
  97 (1988) 651--662.

\bibitem{Coy94}
A.~Coy, P.~T. Callaghan, Pulsed gradient spin echo nuclear magnetic resonance
  for molecules diffusing between partially reflecting rectangular barriers, J.
  Chem. Phys. 101 (1994) 4599--4609.

\bibitem{Grebenkov14}
D.~S. Grebenkov, D.~V. Nguyen, J.-R. Li, Exploring diffusion across permeable
  barriers at high gradients. i. narrow pulse approximation., J. Magn. Reson.
  248 (2014) 153--163.

\bibitem{Pontrelli07}
G.~Pontrelli, F.~de~Monte, Mass diffusion through two-layer porous media: an
  application to the drug-eluting stent, Int. J. Heat Mass Transf. 50 (2007)
  3658--3669.

\bibitem{Todo13}
H.~Todo, T.~Oshizaka, W.~R. Kadhum, K.~Sugibayashi, Mathematical model to
  predict skin concentration after topical application of drugs., Pharmaceutics
  5 (2013) 634--651.

\bibitem{Farago18}
S.~Regev, O.~Farago, Application of underdamped langevin dynamics simulations
  for the study of diffusion from a drug-eluting stent., Phys. A, Stat. Mech.
  Appl. 507 (2020) 231--239.

\bibitem{bressloff-synapse}
P.~C. Bressloff, 2d interfacial diffusion model of inhibitory synaptic receptor
  dynamics, arXiv:2212.05010 (2023).

\bibitem{schumm-synapse}
R.~D. Schumm, P.~C. Bressloff, Local accumulation times in a diffusion-trapping
  model of receptor dynamics at proximal axodendritic synapses, Phys. Rev. E
  105 (2022) 064407.

\bibitem{kk-1}
O.~Kedem, A.~Katchalsky, Thermodynamic analysis of the permeability of
  biological membrane to non-electrolytes, Biochim. Biophys. Acta 27 (1958)
  229--246.

\bibitem{kk-2}
A.~Katchalsky, O.~Kedem, Thermodynamics of flow processes in biological
  systems, Biophys. J. 2 (1962) 53--78.

\bibitem{kk-3}
A.~Kargol, M.~Kargol, S.~Przestalski, The kedem-katchalsky equations as applied
  for describing substance transport across biological membranes, Cell. Mol.
  Biol. Lett. 2 (1996) 117--124.

\bibitem{lejay}
A.~Lejay, The snapping out brownian motion, The Annals of Applied Probability
  26 (2016) 1727--1742.

\bibitem{Lejay18}
A.~Lejay, Monte carlo estimation of the mean residence time in cells surrounded
  by thin layers., Mathematics and Computers in Simulation 143 (2018) 65--77.

\bibitem{lt-1}
P.~L\`evy, Sur certaines processus stochastiques homog\`enes, Compos. Math. 7
  (1939) 283.

\bibitem{lt-2}
S.~N. Majumdar, Brownian functionals in physics and computer science, Curr.
  Sci. 89 (2005) 2076.

\bibitem{lt-3}
H.~P. McKean, Brownian local time, Adv. Math. 15 (1975) 91--111.

\bibitem{bressloff-snob-1d}
P.~C. Bressloff, A probabilistic model of diffusion through a semipermeable
  barrier, Proc. R. Soc. A 478 (2022) 20220615.

\bibitem{bressloff-snob-high-d}
P.~C. Bressloff, Renewal equations for single-particle diffusion through a
  semipermeable interface, Phys. Rev. E 107 (2023) 014110.

\bibitem{3d-narrow-capture}
P.~C. Bressloff, The 3d narrow capture problem for traps with semipermeable
  interfaces, arXiv:2211.12472 (2023).

\bibitem{Farago20}
O.~Farago, Algorithms for brownian dynamics across discontinuities, J. Chem.
  Phys. 423 (2020) 109802.

\bibitem{grebenkov-1}
D.~S. Grebenkov, Probability distribution of the boundary local time of
  reflected brownian motion in euclidean domains, Phys. Rev. E 100 (2019)
  062110.

\bibitem{grebenkov-2}
D.~S. Grebenkov, Paradigm shift in diffusion-mediated surface phenomena, Phys.
  Rev. Lett. 125 (2020) 078102.

\bibitem{bressloff-bf}
P.~C. Bressloff, Diffusion-mediated absorption by partially reaction targets:
  Brownian functionals and generalized propagators, J. Phys. A 55 (2022)
  205001.

\bibitem{sylvain}
M.~Sylvain, T.~Etienne, Monte carlo approximation of the neumann problem, Monte
  Carlo Methods and Applications 19 (2013) 201--236.

\bibitem{wos}
M.~E. Muller, Some continuous monte carlo methods for the dirichlet problem,
  Ann. Math. Statist. 27 (1956) 569--589.

\bibitem{kinetic-monte-carlo}
J.~Cherry, A.~E. Lindsay, A.~Navarro~Hern\'{a}ndez, B.~Quaife, Trapping of
  planar brownian motion: Full first passage time distributions by kinetic
  monte carlo, asymptotic, and boundary integral methods, Multiscale Modeling
  \& Simulation 20~(4) (2022) 1284--1314.

\bibitem{elton}
P.~Hsu, Reflecting brownian motion, boundary local times and the neumann
  problem, Dissertation Abstracts International Part B: Science and Engineering
  45 (1984).

\bibitem{Ward15}
V.~Kurella, J.~C. Tzou, D.~Coombs, M.~J. Ward, Asymptotic analysis of first
  passage time problems inspired by ecology., Bull Math Biol. 77 (2015)
  83--125.

\bibitem{narrow-capture-dirichlet}
A.~E. Lindsay, R.~T. Spoonmore, J.~C. Tzou, Hybrid asymptotic-numerical
  approach for estimating first passage time densities of the two-dimensional
  narrow capture problem, Phys. Rev. E 94 (2016) 042418.

\bibitem{narrow-capture-robin}
P.~C. Bressloff, Narrow capture problem: An encounter-based approach to
  partially reactive targets, Phys. Rev. E 105 (2022) 034141.

\bibitem{narrow-capture-reset}
P.~C. Bressloff, R.~Schumm, The narrow capture problem with partially absorbing
  targets and stochastic resetting, Multiscale Model. Simul. 20 (2022) 101137.

\bibitem{narrow-capture-flux}
P.~C. Bressloff, Asymptotic analysis of extended two-dimensional narrow capture
  problems, Proc. Roy. Soc. A 477 (2021) 20200771.

\bibitem{sde-pde}
E.~Pardoux, A.~Rascanu, Stochastic Differential Equations, Backward SDEs,
  Partial Differential Equations, Springer, New York, 2014.

\bibitem{ward-gf}
A.~Cheviakov, M.~Ward, Optimizing the principal eigenvalue of the laplacian in
  a sphere with interior traps, Math. Comp. Modeling 53 (2011) 042118.

\end{thebibliography}
\end{document}